\documentclass[12pt,draftcls,onecolumn]{IEEEtran}
\usepackage{latexsym, amsmath,amsfonts}
\usepackage{amssymb}
\usepackage{boxedminipage}
\usepackage{graphicx}
\usepackage{multicol}
\usepackage{epstopdf}

\newtheorem{theorem}{Theorem}[section]

\newtheorem{lemma}{Lemma}[section]

\newtheorem{remark}{Remark}[section]

\newtheorem{algthm}{Algorithm}[section]

\newtheorem{assumption}{Assumption}%[section]
\def\EE{{\mathbb E}}

\def\PP{{\mathbb P}}
\def\RR{{\mathbb R}}
\def\BFK{{\bf K}}

\def\dd{{\mathrm d}}
\def\ff{{\mathcal F}}
\def\LL{{\mathcal L}}
\def\NN{{\mathcal N}}

\def\a{\alpha}
\def\bt{\beta}
\def\e{\varepsilon}

\def\kp{\kappa}
\def\lmd{\lambda}
\def\sgm{\sigma}
\def\sgmaste{{\sigma^\ast_e}}

\def\tht{\theta}
\def\thtast{{\theta^\ast}}
\def\thtastA{{\theta^\ast_A}}
\def\thtastB{{\theta^\ast_B}}
\def\thtastC{{\theta^\ast_C}}
\def\thtastD{{\theta^\ast_D}}
\def\thtastF{{\theta^\ast_F}}

\def\Del{\Delta}

\def\Om{\Omega}
\def\om{\omega}

\def\bA{\bar{A}}

\def\bC{\bar{C}}
\def\be{\bar{\varepsilon}}
\def\bF{\bar{F}}
\def\bG{\bar{G}}
\def\bH{\bar{H}}

\def\bs{\bar{s}}
\def\bt{\bar{t}}

\def\bX{\bar{X}}

\def\hPhi{\widehat{\Phi}}

\def\hr{r}

\def\hy{\hat{y}}
\def\ole{\overline{\varepsilon}}

\def\tC{\widetilde{C}}

\def\te{\tilde{\varepsilon}}
\def\tie{\tilde{e}}
\def\tF{\widetilde{F}}

\def\tI{\widetilde{I}}

\def\tu{\widetilde{u}}

\def\tv{\widetilde{v}}

\def\tw{\widetilde{w}}

\def\ty{\widetilde{y}}

\begin{document}
%
% note title
\title{Adaptive input design for LTI systems$^\star$
\thanks{ $\star$ This work was supported by
the European Research Council under the advanced grant
LEARN, contract 26738, and by the Swedish Research Council under contract 621-2009-4017. }
}

\author{ Lirong Huang, H{\aa}kan Hjalmarsson,  L{\'a}szl{\'o}  Gerencs{\'e}r   % <-this % stops a space
%\thanks{This work was not supported by any organization}% <-this % stops a space
%\thanks{Manuscript received \_\_\_\_\_\_}
\thanks{ Lirong Huang was with ACCESS Linnaeus Center and Automatic Control Lab, KTH Royal Institute of Technology, Stockholm, Sweden and is now with Institute of Molecular Systems Biology, ETH Zurich, Switzerland
  (Email address: lirong.huang@imsb.biol.ethz.ch).}
\thanks{ H{\aa}kan Hjalmarsson is with ACCESS Linnaeus Center and Automatic Control Lab, KTH Royal Institute of Technology, Stockholm, Sweden (Email address: hakan.hjalmarsson@ee.kth.se).}
\thanks{L{\'a}szl{\'o}  Gerencs{\'e}r  is with Institute for Control and Computer Science of the Hungarian Academy of Sciences (MTA SZTAKI), Budapest, Hungary (Email address:
         gerencser.laszlo@sztaki.mta.hu).}
}

%
% The note headers
%\markboth{IEEE TRANSACTIONS ON AUTOMATIC CONTROL, ~Vol.~1,
%No.~11,~January ~2009}{Huang  On stability of HSDSs}
% *** Note that you probably will NOT want to include the author's name in ***
% *** the headers of peer review papers.                                   ***

% If you want to put a publisher's ID mark on the page
% (can leave text blank if you just want to see how the
% text height on the first page will be reduced by IEEE)
%\pubid{0000--0000/00\$00.00~\copyright~2002 IEEE}

%\specialpapernotice{(Invited Note)}

\maketitle

\begin{abstract}
Optimal input design for parameter estimation has obtained
extensive coverage in the past. A key problem here is that the optimal
input depends on some unknown  system parameters
that are to be identified.
Adaptive design is one of the fundamental routes to handle this
problem. Although there exist a rich collection of results on this
problem, there are few results that address
dynamical systems. This paper presents sufficient conditions for
convergence/consistency and asymptotic optimality for a class of
adaptive systems consisting of a recursive prediction error estimator
and an input generator depending on the time-varying parameter
estimates. The results apply to a general family of single input
single output linear time-invariant systems. An important application
is adaptive input design for which the results imply that,
asymptotically in the sample size, an adaptive scheme recovers the
same accuracy as the off-line prediction error method that uses data from an experiment where perfect knowledge of the system has been used to design an optimal input spectrum.
\end{abstract}

%\begin{keywords}
%ARMAX identification, $L$-mixing, quasi-stationarity, recursive estimation,
%stochastic approximation.
%%\PACS
%\end{keywords}

%%%%%%%%%%%%%%%%%%%%%%%%%%%%%%%%%%%%%%%%%%%%%%%%%%%%%%%%%%%%%%%%%%%%%%%%%%%%%%%%
\section{Introduction}
With the rapid developments in model based engineering, compare with the
petrochemical industry where it is reported that all plants employ model
predictive control, the high cost of modeling is coming more and more into
focus as a limiting factor \cite{Zhu:09}. Often the only practical means to modeling is
data-driven modeling, i.e. system identification. For this type of
modeling, the major part of the cost is associated with performing experiments on the plant
in question. A key variable here is the duration of the experiments since it
strongly couples to costs in terms of personell, energy, material and production losses.

For dynamical systems it has been shown that careful design of the experiment
can lead to quite drastic reduction in the required experimental time
as compared to standard white noise excitation  or step testing
\cite{Barenthin&Jansson&Hjalmarsson:05a,Zhu:13}.
It has also been stressed that the experimental
conditions are essential for making system identification robust with
respect to many of the design variables that are involved, e.g. model structure and
orders, and with respect to the resulting end
performance \cite{Hjalmarsson:09}.

The aforementioned observations have prompted renewed interest in optimal experiment
design -- a topic that has been studied extensively over the past century, see, e.g.,
\cite{atkinson01}, \cite{bandara09}, \cite{bernaerts05}, \cite{dror06}, \cite{hahn11}, \cite{lohmann92},
\cite{pronzato08}, \cite{wynn70}  and references therein. Recent
advances include novel computationally tractable algorithms
\cite{jansson05}, least-costly and application oriented frameworks
\cite{Geversetal:06a,Hjalmarsson:09}, closed-loop methods
\cite{Hjalmarsson&Jansson:06,Hildebrand:15,Rathousky:13,Marafioti:13,Larsson:13b},
and extensions to non-linear models \cite{Valenzuela:13a,Forgione:14a,DeCock:13}.

A key problem in optimal experiment design is that the optimal
experiment typically depends on the system parameters that are to be identified. One of the fundamental routes
to cope with this problem is to employ adaptive schemes,
meaning that
as information from the system is gathered the experimental conditions are
changed. Adaptive design is usually called sequential design in the
statistics literature, where there exist a rich collection of results
and applications (see, e.g., \cite{lai01} and the references
therein).

When only the input excitation is considered part of the
experiment design, we will use the terminology input design.
Adaptive input design has been studied in many
works in engineering literature (see, e.g., \cite{lindqvist01},
\cite{perez01}, \cite{sasena02}, \cite{gerencser05},
\cite{gerencser09} and \cite{huang12}). However, as pointed out in
\cite{hjalmarsson05} and \cite{gerencser09}, there are few results
that address this problem for dynamical systems. Given the increasing practical
relevance of input design, it is becoming urgent to provide a
solid theoretical foundation for such methods.

When the system is linear time-invariant and belongs to the model set,
and the input is (quasi-)stationary, it is only the second order
properties of the input that asymptotically (in the sample size)
influence the model quality. Thus in this case it is the
spectrum, or equivalently the autocorrelation sequence, of the input that
is the design variable in optimal input design. The actual
input sequence can be generated by  filtering white noise through an
input spectrum shaping filter corresponding to a stable spectral factor of the optimal input
spectrum \cite{jansson05}. Building on this, an obvious approach to
adaptive input design is to combine a recursive identification
scheme with a time-varying input spectrum shaping filter, computed
from the solution of the optimal input design problem using the the
most recent model estimate as a substitute for the true system.

Such a certainty equivalence approach leads to an adaptive feedback
system where, similar to adaptive control, the input properties change
over time depending on the response of the system.
%A major difference as compared to adaptive control
%is that when the system to be identified is open loop stable (or when
%a stabilizing controller is known), stability of the overall adaptive
%system can be easily ensured by enforcing the input to be bounded in the algorithm.
%However,
From a performance perspective there are several issues that
are non-trivial to analyze:
\begin{itemize}
  \item[(i)] Under which conditions will the
    parameter estimates of such a procedure converge?

\item[(ii)] If the algorithm
converges, will it be consistent, i.e. will the model parameters
correspond to the true system parameters?

\item[(iii)] If the algorithm converges to a correct system description,
how does the resulting (large sample) accuracy compare to the accuracy
an oracle, having access to the unknown true parameters for the
experiment design already at the beginning of the experiment, could
achieve?
\end{itemize}
In regards to (iii), notice that even if the parameters
converge to the true values so that, as the experiment time progresses
towards infinity, the input behaves closer and closer to a stationary signal having the
optimal spectrum, suboptimal experimental conditions prevail in
the meantime and it is not evident that the algorithm is able to catch
up with the loss of accuracy this causes -- this strongly depends on the
rate of convergence of the algorithm.

An early version of the above concept was presented in
\cite{lindqvist01}. A severe limitation was that the parameter estimation was not recursive,
requiring re-identification using all past data for each new
measurement. Furthermore, no statistical analysis was provided and
even if, for this off-line algorithm, (i) and (ii) can be dealt with rather
straightforwardly using results from \cite{Ljung78}, (iii) is
non-trivial to analyze since the input signal is non-stationary.

Subsequently, the recursive certainty equivalence approach adopted in
this contribution was
outlined in \cite{gerencser05}, but without formal treatment of
(i)--(iii). Recently, \cite{gerencser09} takes a different approach
and focus on a smaller class of problems, namely, identification of
ARX systems with input filter of finite impulse response (FIR) type as
in \cite{lindqvist01}. The advantage of using ARX-models is that the
analysis of the recursive least-squares  method can be carried out
with a powerful result in \cite{lai82}.

There exists an extensive body of literature on general recursive stochastic
algorithms,
e.g. \cite{ljung77b,kushner03,ljung83,chen02,chen91,gerencser92,gerencser06,chen10}.
Building on this work, the objective of this paper is
to strengthen the theoretical foundations of the adaptive
input design framework outlined in
\cite{gerencser05}, providing results for (i)--(iii),
hereby validating current practice in input design.

While we will cover (i) and (ii), our primary objective will be to
deal with (iii). In particular, with $\theta^\ast$, $\theta_n$ and
$\theta_n^\ast$ denoting the true parameter vector, the parameter estimate
in the adaptive algorithm, and the off-line parameter estimate obtained from an
experiment using the optimal input, respectively,
we will be interested in establishing
conditions for when adaptive input design asymptotically yields
the same asymptotic accuracy as the optimal non-adaptive design in the
sense that
\begin{align*}
\sqrt{n}(\theta_n-\theta^\ast)
\end{align*}
and
\begin{align*}
\sqrt{n}(\theta_n^\ast-\theta^\ast)
\end{align*}
have the same asymptotic distribution. A pre-requisite for this is (of course)
that the recursive estimation algorithm is able to achieve
this when the optimal input is used. Another ambition has been to
cover the general class of single-input single-output (SISO) linear
time-invariant (LTI) systems and associated model structures
considered in \cite{ljung99}. The recursive prediction error (RPE) approach
\cite{ljung77b,ljung83} fulfills these objectives. However, this
algorithm requires a projection mechanism and one generally cannot
exclude the possibility that the sequence of estimates gets trapped at
the boundary where the projection takes place. In the closely related
approach \cite{gerencser92,gerencser06}, the projection is replaced by
a resetting mechanism which allows almost sure convergence to the true parameter
vector to be established. A restrictive assumption here is that the
asymptotic prediction error criterion  is only allowed to have
the true parameter vector as stationary point. This is a more severe
conditition than identifiability. However, for a method that, as in the
case of RPE, is based on gradient based non-linear search the best one can hope for
is that convergence takes place to the set of stationary points.
%In this respect,
%\cite{ljung77b,ljung83} provide the more general result that
Notice that
the corresponding off-line result \cite{Ljung78}, which proves
convergence to the global minimum, makes the assumption that the
global minimum can be found - something that is not easy to guarantee
in practice using gradient based methods, being on-line or off-line.
However, as
our focus is (iii), which has convergence to the true system
parameters as a pre-requisite, we have chosen to base our algorithm
and analysis on the work \cite{gerencser92,gerencser06}, thus avoiding the issue of clustering
at the boundary. Recently, a novel recursive algorithm for ARMAX
models has been proposed in \cite{chen10} for which a powerful
convergence result has been established. Unfortunately, for our
considerations, this convergence result applies only when the input is white and,
furthermore, the asymptotic accuracy of this algorithm is not known, and hence, at least at present, this
algorithm is not suited to our purpose.

The paper starts off in Section \ref{sec-SysInput} by introducing the
system and model assumptions, together with the input signal
generation mechanism that will be employed. The latter depends on the
estimated parameter vector. Prediction error
identification is discussed in Section \ref{sec-PEMestimators},
leading up to the presentation of the complete adaptive algorithm,
comprising the true system, the recursive estimation algorithm and the
input generator, at the end of the section. Formal results on
convergence/consistency and asymptotic distribution
for the adaptive system are provided in Section
\ref{sec-convergence}. These results are quite general in that they
make no specific use of the functional relationship between the
parameter estimate and the input generator, other than that this is a sufficiently smooth map.
These results are then  placed in the context of adaptive input design in the following
Section \ref{sec-AID}, where a complete adaptive input design
algorithm is presented, together with the result that this algorithm
achieves the same asymptotic accuracy as an oracle. The algorithm is
illustrated on a numerical example in Section
\ref{sec-numerical}. Conclusions are provided in Section
\ref{sec:concl}. Proofs are provided in the appendices.

\noindent {\it Notation}: Throughout the paper, unless otherwise specified, we will employ the
following notation.
Our problem will be embedded in an underlying complete probability space $(\Om, \ff,
\PP)$, where $\Om$ is the sample space, $\ff$ is the $\sigma$-algebra that defines events $E$ in $\Om$ which are measurable, i.e., for which the probability $\PP (E)$ is defined. Let $\EE [\cdot]$ be the expectation operator with respect to the probability measure.  If $A$ is a vector or matrix, its
transpose is denoted by $A^{T}$. If $P$ is a square matrix, $P>0$
($P<0$) means that $P$ is a symmetric positive (negative) definite
matrix of appropriate dimensions while $P \ge 0$ ($P \le 0$) is a
symmetric positive (negative) semidefinite matrix. If the square matrix $P$ is nonsingular, its inverse is denoted by $P^{-1}$. $I_m$ stands for
the identity matrix of order $m$, $0_{m \times n}$ stands for the zero
matrix of dimensions $m \times n$, $0_m = 0_{m \times 1}$ stands for
the zero vector of dimension $m$, and $0$
denotes the zero matrix of appropriate dimensions.
%For simplicity, we also denote by $0$ the zero matrix of appropriate dimensions where there is no ambiguity.
Denote by $\lambda_M
(\cdot)$, $\lambda_m (\cdot)$ and $\rho(\cdot)$ the maximum
eigenvalue, minimum eigenvalue and spectral radius
of a matrix, respectively.
 For a vector, let $| \cdot |$ denote the Euclidean norm and for a
 matrix the norm induced by the Euclidean norm. Unless explicitly
stated, matrices are assumed to have real entries and compatible
dimensions.

%%%%%%%%%%%%%%%%%%%%%%%%%%%%%%%%%%%%%%%%%%%%%%%%%%%%%%%%%%%%%%%%%%%%%%%%%%%%%%%%
\section{LTI system and input signal} \label{sec-SysInput}

%\subsection{The system}

Let us consider a general form of SISO LTI models (see, e.g., \cite{ljung99})
\begin{equation} \label{system-true}
A(q,\tht_A) y_n = \frac{ B(q,\tht_B)} { F(q,\tht_F)} u_{n} + \frac{ C (q,\tht_C)} { D (q,\tht_D)} e_n
\end{equation}
where $A(q,\tht_A)$, $B(q,\tht_B)$, $C(q,\tht_C)$, $D(q,\tht_D)$ and $F(q,\tht_F)$ are polynomials in the backward shift operator $q^{-1}$ of degrees $p_a$, $p_b$, $p_c$, $p_d$ and $p_f $, respectively,
\begin{eqnarray}  \label{dfn-ABCDF-theta}
&& A (q, \tht_A) = 1 + \sum_{j=1}^{p_a} a_j q^{-j}, \quad B (q, \tht_B) = \sum_{j=1}^{p_b} b_j q^{-j}, \quad C (q, \tht_C) = 1 + \sum_{j=1}^{p_c} c_j q^{-j},
   \nonumber \\
&&    D (q, \tht_D) = 1 + \sum_{j=1}^{p_d} d_j q^{-j},  \quad
   F (q, \tht_F) = 1 + \sum_{j=1}^{p_f} f_j q^{-j},
\end{eqnarray}
where the parameters to be estimated are
$\tht_A = \left( a_1 \,\,\, a_2 \,\,\, \cdots \,\,\,  a_{p_a} \right)^T $, $\tht_B = \left( b_1 \,\,\, b_2 \,\,\, \cdots \,\,\,  b_{p_b} \right)^T $,  $\tht_C = \left( c_1 \,\,\, c_2 \,\,\, \cdots \,\,\,  c_{p_c} \right)^T $, $\tht_D = \left( d_1 \,\,\, d_2 \,\,\, \cdots \,\,\,  d_{p_d} \right)^T $ and $\tht_F = \left( f_1 \,\,\, f_2 \,\,\, \cdots \,\,\,  f_{p_f} \right)^T $. We collect all parameters into $\tht  = \begin{bmatrix} \tht_A^T & \tht_B^T & \tht_F^T   & \tht_C^T  & \tht_D^T \end{bmatrix}^T \in\RR^{p_\tht}$ where $ p_{\tht} =p_a +p_b + p_f + p_c + p_d \ge 1$. We will assume that $\theta\in D_\tht \subset \RR^{p_{\tht}}$, where the set $D_\tht$ will be specified below.

The model (\ref{system-true}) is very general, allowing the dynamics from input $u$ to output $y$ to be modeled separately from the measurement noise, but also allowing for the input and noise to share dynamics.  We will make the following assumptions on the system.
\begin{assumption}  \label{assumption-stable}
  The true system is given by  (\ref{system-true}) for some $\tht$.
Labeling true parameters with asterisks, e.g. $\tht^\ast$, $\tht_A^\ast$, and $a_1^\ast$, and the true polynomials by $A^\ast (q)=A (q, \thtastA)$, etc, it holds that
 $A^\ast (z) \ne 0$, $F^\ast (z) \ne 0$, $C^\ast (z) \ne 0$ and $D^\ast (z) \ne 0$ for all $ |z| \ge 1$, and also that $\tht^\ast\in {\rm int} D_\tht$.
Furthermore, the system is at rest prior to time $n=0$, i.e., $y_n = u_n = e_n =0$ for $n< 0$.
\end{assumption}
The conditions in Assumption \ref{assumption-stable} on $C^\ast (z)$ and $D^\ast (z) $ are not restrictive \cite{ljung99},  while those on $A^\ast (z)$ and $F^\ast (z)$  impose stability of the system.
Let a minimal state-space representation of the true system be given by
\begin{equation}  \label{system-ss}
\begin{array}{rcl}
\xi_{n+1} &=& A_\xi \xi_n + B_\xi u_n + K_\xi e_n \\
y_n &=& C_\xi \xi_n + e_n \end{array}
\end{equation}
where $\xi_n \in \RR^{n_\xi}$, $u_n \in \RR$ and $y_n \in \RR$ represent the states, input and output of the system, respectively. Assumption \eqref{assumption-stable} implies that the transition
matrix $A_\xi$ has all its eigenvalues strictly inside the unit circle, i.e., the system (\ref{system-ss}) is internally stable, and that the matrix
$A_\xi - K_\xi C_\xi $ has all its eigenvalues strictly inside the unit circle, i.e., the system (\ref{system-ss}) is inversely stable from $\{ y_n \}$ to $\{ e_n \}$. For the noise process we have the following assumption.
\begin{assumption}  \label{assumption-noise-e}
The noise process $ \{ e_n \}$ is a sequence of independent random variables such that
\begin{equation}   \label{dfn-en}
\EE [ e_n ] =0, \quad \EE [ e_n^2 ]= \sgmaste^2, \quad \sup_n \EE [ \exp{(\a_e e_n^2 )} ] < \infty
\end{equation}
 for some $\a_e  >0$,  where  $\sgmaste^2 >0 $ is unknown.
\end{assumption}
Assumption \ref{assumption-noise-e} on the noise is certainly satisfied for
  independent and identically distributed (i.i.d.) Gaussian sequences (see \cite{chen91,gerencser92,gerencser06}). We will impose a standard identifiability condition.

\begin{assumption}  \label{assumption-coprime}
The model structure (\ref{system-true}) is globally identifiable at $\thtast$ (see \cite[Theorem 4.1, p116]{ljung99}), i.e.,
\begin{itemize}
\item[i)]  there is no common factor to all $z^{p_a} A^\ast (z)$, $z^{p_b} B^\ast (z)$ and $z^{p_c} C^\ast (z)$,
\item[ii)]  there is no common factor to  $z^{p_b} B^\ast (z)$ and $z^{p_f} F^\ast (z)$,
\item[iii)]  there is no common factor to   $z^{p_c} C^\ast (z)$ and $z^{p_d} D^\ast (z)$,
\item[iv)]  if $p_a \ge 1$, then there must be no common factor to  $z^{p_f} F^\ast (z)$ and $z^{p_d} D^\ast (z)$,
\item[v)]  if $p_d \ge 1$, then there must be no common factor to  $z^{p_a} A^\ast (z)$ and $z^{p_b} B^\ast (z)$,
\item[vi)]  if $p_f \ge 1$, then there must be no common factor to  $z^{p_a} A^\ast (z)$ and $z^{p_c} C^\ast (z)$;
\end{itemize}
\end{assumption}
%and
The convex set $D_\tht$ to which $\theta$ is restricted is in this paper defined as
\begin{equation} \label{dfn-D_tht}
D_\tht =\{ \tht: g( \tht ) \le 1, \tht_C \in D_C, \tht_F \in D_F \} ,
\end{equation}
where $g: \RR^{p_{\tht}} \to \RR_+$ is a continuous function,
$D_C$ and $D_F$ are  both compact sets corresponding to stable
polynomials, i.e., $C(z, \tht_C) \ne 0$ and $F(z, \tht_F) \ne 0$ for
all $|z| \ge 1$ on $D_\tht$. In fact, we will impose a stricter
condition. For this we introduce the joint spectral radius for a set of bounded matrices $\Sigma$, defined as
\begin{equation}   \label{joint-spectral-radius-CF}
\rho (\Sigma) = \limsup_{ n \to \infty} \rho_n (\Sigma),\;
\text{where } \rho_n (\Sigma) = \sup \{ [ \rho( A )]^{1/n}:  A \in \Sigma^n \}
\end{equation}
where $ \Sigma^n = \{ A_n \cdots A_1: A_k \in \Sigma, k =1,   \cdots, n \} $.
Let $\tC (\tht_C) \in \RR^{p_c \times p_c}$ and $\tF (\tht_F) \in
\RR^{p_f \times p_f}$ be the companion matrices of $C(q, \tht_C)$ and $F(q, \tht_F)$,  that is,
\begin{equation*}
\tC (\tht_C) = \begin{bmatrix} -c_1 & - c_2  & \cdots  & - c_{p_c-1} & - c_{p_c} \\
                                  1   &   0    & \cdots  &  0          &  0        \\
                                  0   &   1    & \cdots  &  0          &  0        \\
                               \vdots & \vdots & \ddots  &  \vdots     & \vdots    \\
                                  0   &   0    & \cdots  &  1          &  0        \end{bmatrix} \,\, {\rm and} \,\,
\tF (\tht_F) = \begin{bmatrix} -f_1 & - f_2  & \cdots  & - f_{p_f-1} & - f_{p_f} \\
                                  1   &   0    & \cdots  &  0          &  0        \\
                                  0   &   1    & \cdots  &  0          &  0        \\
                               \vdots & \vdots & \ddots  &  \vdots     & \vdots    \\
                                  0   &   0    & \cdots  &  1          &  0        \end{bmatrix}.
\end{equation*}
Then we will use the following assumption.
\begin{assumption}
  \label{ass:jointstab}
The sets $D_C$ and $D_F$ are compact and
the joint spectral radii of $\Sigma_C = \{ \tC (\tht_C): \tht_C \in  D_C \}$ and
$\Sigma_F = \{ \tF (\tht_F): \tht_F \in  D_F \}$ are less than one.
Furthermore, the function $g: \RR^{p_{\tht}} \to
\RR_+$ in \eqref{dfn-D_tht} is continuous.
\end{assumption}
A common way, see, e.g., Condition 4.5 in \cite{gerencser92}, to
ensure Assumption \ref{ass:jointstab} is to assume
that there are  positive definite matrices $V_c \in \RR^{ p_c \times p_c}$, $V_f \in \RR^{ p_f \times p_f}$  satisfying
\begin{eqnarray}    \label{dfn-Vc}
&& \tC^T  (\tht_C) V_c \tC  (\tht_C)  <    \lambda V_c ,  \qquad \forall \, \tht_C \in D_C  \\
    \label{dfn-Vf}
&& \tF^T  (\tht_F) V_f \tF  (\tht_F)  <   \lambda V_f ,  \qquad \forall \, \tht_F \in D_F
\end{eqnarray}
for some $0<\lambda<1$, respectively.

\begin{remark} \label{Remark-VcVf}
Suppose that $ D_C$ and $ D_F $ are convex polyhedra with vertices $\tht_{C,k}$, $k=1, \cdots, n_{c}$, and $\tht_{F,j}$, $j=1, \cdots, n_{f}$, respectively.
The  positive definite matrices $V_c$ and $V_f$ satisfy (\ref{dfn-Vc}) and (\ref{dfn-Vf}) for all $\tht_C \in D_C$ and $\tht_F \in D_F$
if the following linear matrix inequalities (LMIs) %%%(see \cite{boyd94})
\begin{eqnarray*} % \label{commonLya-DC}
   \tC_k^T  V_c \tC_k < \lambda V_c, \quad k=1, \cdots, n_{c}, \quad {\rm and}\quad
   %\label{commonLya-DF}
   \tF_j^T   V_f \tF_j < \lambda V_f, \quad j=1, \cdots, n_{f},
\end{eqnarray*}
hold, respectively, where $\tC_k =\tC  (\tht_{C,k})$ and $\tF_j =\tF  (\tht_{F,j})$.  This can be easily shown by using that $[ \tC_1^T V_c \tC_2 + \tC_2^T V_c \tC_1] \le [ \tC_1^T V_c \tC_1 + \tC_2^T V_c \tC_2] $ and hence
\begin{eqnarray*}
   [ a \tC_1 + b \tC_2 ]^T V_c  [ a \tC_1 + b \tC_2  ]
   < (a^2+ b^2) \lambda V_c +  ab  [ \tC_1^T V_c \tC_1 + \tC_2^T V_c \tC_2] < (a^2+2ab+ b^2) \lambda V_c  =\lambda V_c
\end{eqnarray*}
for all $a\ge 0$, $b\ge 0$ such that $a+b=1$.
%%%%This set of LMIs  (\ref{commonLya-DC}) and (\ref{commonLya-DF}) can be easily solved with some toolboxes, e.g., the Matlab LMI toolbox.
\end{remark}

\begin{remark}  \label{Remark-ARARX2}
Assumption \ref{ass:jointstab} is certainly restrictive, but notice that
Conditions (\ref{dfn-Vc})-(\ref{dfn-Vf}) represent the state-of-the
art in recursive parameter estimation (see the remark below Condition
4.5 in \cite{gerencser92}).
\end{remark}

\begin{remark}  \label{Remark-ARARX}
Observe that Assumption \ref{ass:jointstab} is trivially satisifed for ARARX systems, i.e., model (\ref{system-true}) with $p_c =p_f =0$. The ARARX is a stochastic model commonly used in  economics, engineering, health and medical science literature
(see, e.g., \cite{almeida04,diversi10,haest90,kohn79,nollo01,nooraii99,noriega12,porta98,tjarnstrom02} and the references therein). As an application example of  our proposed method, a problem of adaptive input design for a class of ARARX models will be considered in Section \ref{sec-numerical}.

For ARMAX systems, i.e. when $p_d=p_f=0$, Assumption \ref{ass:jointstab} can be relaxed when
the input is not adaptively
updated, e.g. the method in \cite{chen10} applies to white inputs. See also \cite{huang14}.
%Notice that condition (\ref{dfn-Vc}) is not employed by  some results, see  and \cite{huang14}. It is our future work to develop the underlying theory and remove these restrictive conditions.
\end{remark}

\begin{remark} \label{remark-globalID}
By  continuity of the model structure (\ref{system-true}) (see also Appendix A), there exists a compact subset $D_{\tht 0} \subseteq D_\tht$ with $\thtast \in {\rm int} D_{\tht 0}$ such that Assumption \ref{assumption-coprime} holds for all $\tht \in D_{\tht 0} $, or say, the model structure (\ref{system-true}) is globally identifiable at all $\tht \in D_{\tht 0} $, where $D_\tht$ is given by (\ref{dfn-D_tht}).
\end{remark}

The input signal $\{ u_n \}$ is defined in terms of an external source represented by a state-space system that is at rest prior to
time $n=0$,
\begin{equation}  \label{input-ss}
\begin{array}{rcl}
z_{n+1} & = & A_z (r(\tht_{n})) z_n + B_z (r(\tht_{n})) s_n,\, n\geq
0,\quad z_{n}=0,\, n<0  \\
u_n & = & C_z (r(\tht_{n})) z_n + D_z (r(\tht_{n})) s_n.
\end{array}
\end{equation}
Here $\tht_n$ is the estimate of $\thtast$ and
the state-space matrices $A_z \in \RR^{m \times m}$, $B_z \in \RR^{m}$, $C_z^T \in \RR^{m }$, $D_z \in \RR$, with $m$ being a finite non-negative integer, are functions of the variable $r$, which in turn is a function of the model parameters, i.e. $r:\;\RR^{p_\theta}\rightarrow \RR^{p_r}$.

We will need bounded-input bounded-output (BIBO) stability of the
input signal generator \eqref{input-ss}. According to \cite[Corollary 1.1, p21]{jungers09} (see also \cite{daubechies92} and \cite{berger92}) and Lemma 27.4 in \cite{rugh96}, the time-varying system (\ref{input-ss}) is BIBO stable the following assumption holds.
\begin{assumption}  \label{assumption-input-signal}
  The set of matrices $\{A_z(r(\theta)),\; B_z(r(\theta)),\;
  C_z(r(\theta)),\; D_z(r(\theta)):\theta\in D_\theta\}$ is bounded,
  and the joint spectral radius of $\Sigma_z = \{ A_z (r(\theta)):
  \tht \in D_\tht  \}$ is less than one.

Furthermore, the process $\{ s_n \}$  is a sequence of independent random variables, independent of $\{ e_n \}$,
such that
\begin{equation}  \label{dfn-sn}
 \EE [ s_n ]=0 , \;\;\; \EE [ s_n^2 ] =1, \;\;\;
\sup_n \EE [ \exp{(\a_s s_n^2 )} ] < \infty
\end{equation}
for some $\a_s >0$.
\end{assumption}
%\begin{assumption}  \label{input-joint-spectral-radius}
%\end{assumption}
Since the input generator (\ref{input-ss}) is in the hands of the user, Assumption
\ref{assumption-input-signal} can be ensured by appropriate
design using techniques from the theory on stability of linear
time-varying systems, see, e.g., the discussion after Assumption \ref{ass:jointstab}.
As we will see in Section \ref{sec-AID}, in adaptive input design $A_z(\cdot)$
does typically not depend on $\theta$, in which case the condition on
the joint spectral radius of $\Sigma_z$ is trivially satisfied.

%%%%%%%%%%%%%%%  Section Prediction error estimation
\section{Prediction error estimation}  \label{sec-PEMestimators}
%\subsection{Consistency}
%Let $\tht \in {\rm int} D_\tht$ and $\sgm^2 \in  D_\sgm$.
For any  $\tht  \in  D_\tht$, define the prediction error process  by
\begin{equation}  \label{error-frozen}
\overline{\e}_n (\tht ) =y_n  - \hy_n (\tht)
\end{equation}
for all $n \ge 0$, where $y_n $ is the output of the true system
(i.e. (\ref{system-true}) with $\theta=\theta^\ast$) with a persistently exciting input signal $u_n=u_n(\thtast) $ and $\hy_n (\tht)$ is the one-step predictor for the LTI model (\ref{system-true})
\begin{equation}    \label{predictor}
\hy_n (\tht) = \frac{ D (q, \tht_D) B (q, \tht_B) }{ C (q, \tht_C) F(q, \tht_B)  } u_n
+ \left[ 1 - \frac{ D (q, \tht_D) A (q, \tht_A)  }{ C (q, \tht_C) }  \right] y_n,
\end{equation}
which can also be written as a recursion
\begin{equation}    \label{predictor-recursion}
C (q, \tht_C) F(q, \tht_B) \hy_n (\tht) =  F(q, \tht_B) \left[ C (q, \tht_C) -  D (q, \tht_D) A (q, \tht_A)  \right] y_n
+ D (q, \tht_D) B (q, \tht_B)u_n .
\end{equation}
Introducing the auxiliary variables
\begin{equation}   \label{dfn-w}
w_n (\tht ) = \frac{ B (q, \tht_B)} { F (q, \tht_F)}u_n  \quad \Rightarrow \quad
w_n (\tht ) = \sum_{j=1}^{p_b} b_j u_{n-j}    - \sum_{j=1}^{p_f} f_j w_{n-j} (\tht )
\end{equation}
and
\begin{equation}   \label{dfn-v}
v_n (\tht ) = A (q, \tht_A) y_n- w_n (\tht ) \quad \Rightarrow \quad
v_n (\tht ) = y_n + \sum_{j=1}^{p_a} a_j y_{n-j}  - w_n (\tht ),
\end{equation}
 we have  (see, e.g., \cite{ljung83} and \cite{ljung99})
\begin{equation}  \label{error-recursion}
 \overline{\e}_n (\tht  ) = y_n  - \hy_n (\tht) = y_n - \tht^T \varphi_{n-1} (\tht)
\end{equation}
for all $n$, where $ \varphi_{n}$ is a function of $\theta$ defined by
\begin{equation}  \label{dfn-varphi}
  \varphi_{n} (\tht ) =\begin{bmatrix} -\ty_{n-1}^T
 & \tu_{n-1}^T   & - \tw_{n-1}^T(\tht ) & \te_{n-1}^T(\tht )  & - \tv_{n-1}^T(\tht )\end{bmatrix}^T
\end{equation}
with $\ty_{n-1}  = [ y_{n-1} \, \cdots \, y_{n-p_a}  ]^T \in \RR^{p_a} $, $\tu_{n-1}  = [ u_{n-1}    \, \cdots \, u_{n-p_b}   ]^T \in \RR^{p_b} $,
 $\tw_{n-1}(\tht ) = [ w_{n-1} (\tht ) \, \cdots \,
w_{n-p_f} (\tht )  ]^T \in \RR^{p_f} $, $\te_{n-1}(\tht ) = [ \ole_{n-1} (\tht ) \, \cdots \, \ole_{n-p_c} (\tht ) ]^T \in \RR^{p_c} $,
 $\tv_{n-1}(\tht ) = [ v_{n-1} (\tht ) \, \cdots \, v_{n-p_d} (\tht ) ]^T \in \RR^{p_d} $.

%In this case,
%the input signal $\{ u_n \}$ is generated by system (\ref{input-ss})
%with  fixed $\tht \in D_\tht$ and $r = r (\tht)$.

 The asymptotic cost function
is defined by (see \cite{ljung79} and \cite{gerencser92})
\begin{equation}  \label{dfn-asym-cost-funct}
W (\tht ) = \lim_{ n \to \infty} \frac{1}{2} \EE [ \ole_n^2 (\tht) ].
\end{equation}
Then the gradient and the Hessian of $W$ are given by
\begin{eqnarray}  \label{gradient}
W_\tht (\tht ) & = &\frac{ \partial }{\partial \tht} W(\tht) = \lim_{ n \to \infty} \EE [ \ole_{\tht, n}  (\tht) \ole_n (\tht) ], \\
  \label{Hessian}
  W_{\tht \tht} (\tht ) & = & \frac{ \partial^2 }{\partial  \tht^2} W(\tht)=  \frac{ \partial^2 }{\partial \tht \partial \tht^T} W(\tht)
\end{eqnarray}
respectively, where $\overline{\e}_{\tht,n} (\tht)  =\ole_{\tht, n} (\tht ) = \frac{ \partial }{\partial \tht} \ole_{ n} (\tht ) = [ \frac{ \partial }{\partial \tht_1} \ole_{ n} (\tht ) \,\,\, \cdots \,\,\, \frac{ \partial }{\partial \tht_{p_\tht }} \ole_{ n} (\tht )  ]^T $. Define
\begin{equation}  \label{GHessian}
 G(\tht) =\lim_{ n \to \infty} \EE [ \ole_{\tht, n} (\tht) \ole_{\tht, n}^T (\tht) ]
\end{equation}
for all $\tht \in  D_\tht$.  Then $ W_{\tht \tht} (\thtast ) =G(\thtast) $.

  Note that $y_n =0$, $u_n =0$, $\ole_n =0$ and  $\ole_{\tht, n}=0 $ for all
$n<0$ since the system is at rest prior to time $n=0$.
Let $(\ff_n , \ff_n^+)$, $n \ge 0$, be a pair of families of
$\sgm$-algebras such that (i) $\ff_n \subset \ff$ is monotone
increasing, (ii) $\ff_n^+ \subset \ff$ is monotone decreasing, and
(iii)  $\ff_n$ and  $\ff_n^+$ are independent for all $n \ge 0$.
In this paper, we set $\ff_n = \sgm \{ e_t, s_t: 0 \le t \le n\}$ and $\ff_n^+ = \sgm \{ e_t, s_t: t \ge n+1 \}$.
For simplicity, we write
$\ole_n  = \overline{\e}_n (\tht ) $, $\ole_{\tht,n}  = \overline{\e}_{\tht,n} (\tht ) $, etc. where there is no ambiguity.  The overline
indicates that (\ref{error-frozen}) is defined as a frozen-parameter process (for fixed $\tht \in  D_\tht$).   Denote by  $\e_{n}$ and $\e_{\tht, n}$   the online estimates of $\ole_{n} $ and $\ole_{\tht, n} $, respectively.

According to the model  (\ref{error-recursion}),  the gradient $ \ole_{\theta, n} $  is given by  (see, e.g.,  \cite{gerencser92, ljung83})
\begin{equation} \label{C-e-derivative}
 C(q, \tht_C) \ole_{\theta, n} ( \tht ) =    - \varphi_n (\tht)  + \Psi_{n} (\tht),
\end{equation}
where
\begin{equation}  \label{def_Psi}
    \Psi_{n} (\tht) =  \begin{bmatrix}  \left[ D(q, \tht_D)-1  \right]  \ty_{n-1}  \\   \tu_{n-1} \\ - \tw_{n-1} (\tht) \\ 0 \\ 0 \end{bmatrix} - D(q, \tht_D ) w_{\tht, n} (\tht)
\end{equation}
 with $ \ty_{n-1}$, $ \tu_{n-1}$,  $  \tw_{n-1} (\tht)$  given by (\ref{error-recursion}) and $w_{\tht, n} =w_{\tht, n}  (\tht)$ is defined by
 \begin{equation}   \label{def_Psi}
  F(q, \tht_F) w_{\tht, n} (\tht)  = \begin{bmatrix} 0 & \tu_{n-1}^T & - \tw_{n-1}^T (\tht) & 0 & 0 \end{bmatrix}^T.
 \end{equation}
 Hence $ \varphi_n (\tht)$,  $ \Psi_{n} (\tht)$, $\ole_{\theta, n} ( \tht )$ are $\ff_{n-1}$-measurable.
That is,
\begin{equation} \label{e-derivative}
 \ole_{\theta, n} ( \tht ) =   - \varphi_n (\tht)  + \Psi_{n} (\tht) - \sum_{j=1}^{p_c} c_j  \ole_{\theta, n-j} ( \tht ).
\end{equation}

 By  (\ref{error-recursion}), we have
\begin{eqnarray}
  \label{error-frozen-1}
 \overline{\e}_n (\tht ) &  = & \ty_{n-1}^T ( \tht_A  - \thtastA) - \tu_{n-1}^T ( \tht_B  - \thtastB) + \tw_{n-1}^T  ( \tht_F  - \thtastF) - \te_{n-1}^T
 (\tht_C-\thtastC )     \nonumber \\
& &  {} + \tv_{n-1}^T  ( \tht_D  - \thtastD)   + \Del \tw_{n-1}^T \thtastF -  \Del \te_{n-1}^T  \thtastC + \Del \tv_{n-1}^T  \thtastD + e_n ,
\end{eqnarray}
where $ \Del \tw_{n-1}= \Del \tw_{n-1} (\tht, \thtast ) =   \tw_{n-1} (\tht ) - \tw_{n-1} (\thtast)$, $ \Del \tv_{n-1}= \Del \tv_{n-1} (\tht, \thtast ) =   \tv_{n-1} (\tht ) - \tv_{n-1} (\thtast)$  and $\Del \te_{n-1} =\Del \te_{n-1} (\tht, \thtast )= \te_{n-1} (\tht ) - \tie_{n-1} $ with $\tie_{n-1} = \te_{n-1} (\thtast) %%%= \te_{n-1} (\thtast)
=[ e_{n-1}  \,\,\, \cdots \,\,\, e_{n-p_c} ]^T$.

The true parameter $\thtast$ is obtained as the solution to the equation
\begin{eqnarray} \label{normal-equation}
  \EE [ \ole_{\tht, n}   (\tht)  \ole_n (\tht) ] & = &  -   \EE [ \varphi_n   (\tht)  \ole_n (\tht) ] +  \EE [  \Psi_n   (\tht)  \ole_n (\tht) ] - \sum_{j=1}^{p_c}  \EE [ c_j \ole_{\tht, n-j}   (\tht)  \ole_n (\tht) ] \nonumber \\
& = &    \EE [  \varphi_n (\tht)   \varphi_n^T  (\tht)    ] (\tht - \thtast) -  \EE [ \varphi_n (\tht)     \Del \tw_{n-1}^T  ]\thtastF
 + \EE [  \varphi_n (\tht)    \Del \te_{n-1}^T   ]\thtastC  \nonumber    \\
&&   {}  -  \EE [  \varphi_n (\tht)     \Del \tv_{n-1}^T   ]\thtastD  - \EE [   \varphi_n (\tht)     e_n]  + \EE [  \Psi_n   (\tht)  \ole_n (\tht) ] - \sum_{j=1}^{p_c}  \EE [ c_j \ole_{\tht, n-j}   (\tht)  \ole_n (\tht) ]\nonumber    \\  %
%\end{eqnarray}
%\begin{eqnarray}
& = &  \EE [  \varphi_n (\tht)   \varphi_n^T  (\tht)    ] (\tht - \thtast) -  \EE [ \varphi_n (\tht)     \Del \tw_{n-1}^T  ]\thtastF
 + \EE [  \varphi_n (\tht)    \Del \te_{n-1}^T   ]\thtastC  \nonumber    \\
&&   {}  -  \EE [  \varphi_n (\tht)     \Del \tv_{n-1}^T   ]\thtastD  +  \EE [  \Psi_n   (\tht)  \ole_n (\tht) ] - \sum_{j=1}^{p_c}  \EE [ c_j \ole_{\tht, n-j}   (\tht)  \ole_n (\tht) ]  \nonumber    \\
&= & 0  \quad \; {\rm as}\; \, n \to \infty.
\end{eqnarray}
Notice that $  \ole_n (\thtast) =e_n$ is independent of  $ \varphi_n (\thtast)$,  $ \Psi_{n} (\thtast)$, $\ole_{\theta, n} ( \thtast )$ which are  $\ff_{n-1}$-measurable.  Therefore, $\tht =\thtast$ is a solution to equation (\ref{normal-equation}).
We will impose the following assumption.
\begin{assumption} \label{assumption-unique-solution}
$\tht =\thtast$ is the unique solution to the normal equation  (\ref{normal-equation}) on $ D_\tht$.  %%% ${\rm int} D_\tht$.
\end{assumption}
Assumption  \ref{assumption-unique-solution} implies that $\thtast$ is
consistently estimated when the input is generated by (\ref{input-ss})
with $\tht_{n}$ kept fixed in the input generator.
Assumption \ref{assumption-unique-solution} requires that the transfer function $G_u (q, r(\tht)) = C_z
  (r(\tht)) \big[ q I_{m} - A_z (r(\tht)) \big]^{-1} B_z (r(\tht)) +
D_z (r(\tht))$ is not identically zero, in turn implying that the
corresponding input spectrum  (see, e.g., \cite[Theorem 2.2, p.40]{ljung99})
\begin{equation}  \label{dfn-u-spectrum-positivity}
\Phi_u  (e^{i \om},r(\tht)) =\big| G_u (e^{i \om},r(\tht) ) \big|^2 \ge 0\;\forall \omega,
\end{equation}
is not identically zero. In fact, since $G_u$ is finite-dimensional it
can only have a finite number of zeros on the unit circle, it must
hold that
\begin{equation}  \label{dfn-u-spectrum-strictpositivity}
\Phi_{u} ( e^{i \om},r(\theta))  > 0\; \text{for almost all }\om,
\end{equation}
which means that the input signal $\{ u_n \}$ is persistently exciting (see, e.g., \cite[Definition 13.2, p.414]{ljung99}) when the input filter is fix.
It is possible to influence the uniqueness of the solution to
\eqref{normal-equation} by appropriate choice of the input spectrum, see \cite{Eckhard:13a} and references therein. However, the shape of such spectra depend on the unknown $\theta^\ast$. It would be interesting to develop adaptive schemes based on this type of result so that Assumption \ref{assumption-unique-solution} could be relaxed.

%%%%%
%\subsection{Recursive algorithm with resetting}

The model (\ref{error-frozen}) together with the gradient expression (\ref{e-derivative}) immediately suggests a Newton-type recursive prediction error estimate of $\thtast$ as follows (see, e.g., \cite{gerencser05} and \cite{ljung83})
\begin{eqnarray}  \label{rec-algo-nonresetting-1}
\tht_{n+1} & = & \tht_n - \frac{1}{n+1} R_n^{-1} \e_{\tht, n+1} \e_{n+1},  \\ %\quad \tht_0 \in int D_\tht \\
\label{rec-algo-nonresetting-2}
R_{n+1} & = & R_n + \frac{1}{n+1} ( \e_{\tht, n+1} \e_{\tht, n+1}^T - R_n ), %\quad R_0 \in int D_R
\end{eqnarray}
where $\e_{n+1}$ and $\e_{\tht, n+1}$ are  the online estimates of $\ole_{n} $ and $\ole_{\tht, n} $ given by (\ref{error-recursion}) and  (\ref{e-derivative}), respectively.
%In this case,  $\ty_{n-1} $,   $\tu_{n-1} $ and $\te_{n-1} $ are, correspondingly, their online versions respectively, e.g., $\te_{n-1} = [ \e_{n-1}  \,\,\, \cdots \,\,\, \e_{n-p_c}   ]^T $.

In order to ensure that the estimates do not leave their domain of
definition $D_\theta$, and even stay in a bounded domain, recursive estimation schemes such as (\ref{rec-algo-nonresetting-1})-(\ref{rec-algo-nonresetting-2}) typically need to be complemented with either a projection or a resetting mechanism (see \cite{chen02}, \cite{gerencser92}, \cite{gerencser06}, \cite{gerencser09}, \cite{huang11}, \cite{kushner10}, \cite{kushner03} and \cite{ljung77b}). In this work, we consider the  recursive estimation algorithm (\ref{rec-algo-nonresetting-1})-(\ref{rec-algo-nonresetting-2}) with a resetting mechanism, which is part of the entire adaptive system  (\ref{dfn-Dtht})-(\ref{rec-algo-resetting-3}) below.

 Obviously, we have $ G(\tht)  =  \lim_{ n \to \infty} \EE [
   \ole_{\tht, n} (\tht) \ole_{\tht, n}^T (\tht) ] \ge 0 $ for all
 $\tht \in D_\tht$. Under the above assumptions, we have the following
 result ensuring local identifiability:
\begin{lemma}  \label{G-positive-definite}
There is a subset $D_\thtast \subseteq D_\tht$ such that $\thtast \in {\rm int} D_\thtast$ and
%\begin{equation}  \label{GHessian-positive}
$ G(\tht)   >0 $
%\end{equation}
for all $\tht \in D_\thtast$.
\end{lemma}

{\it Proof:} See Appendix A. \\%%%\ref{proof_of_G_positive}.
Lemma \ref{G-positive-definite} implies that
\begin{equation}  \label{dfn-Rast}
R^\ast = G (\thtast) =W_{\tht \tht} (\thtast ) = \lim_{ n \to \infty} \EE [ \ole_{\tht, n} (\thtast) \ole_{\tht, n}^T (\thtast) ] >0 .
\end{equation}
Let $D_R$ be a compact set of symmetric positive definite matrices defined as
$ D_R = \{ P \in \RR^{p_\tht \times p_\tht}: \kp_1 I_{p_\tht} \le P  \le \kp_2 I_{p_\tht}  \}$, denoted by (\ref{dfn-DR}) below,
where $\kp_1$ and $\kp_2$ are sufficiently small and large positive
constants (that will be given by (\ref{ode-Rt-bounds-all-t}) in
Appendix D), respectively.

In summary, the adaptive system consists of the system
\eqref{system-ss}, the input generator \eqref{input-ss}, the on-line
versions of the prediction-error gradient \eqref{e-derivative}, the
prediction-error itself \eqref{error-frozen-1}, the Newton updates
\eqref{rec-algo-nonresetting-1}--\eqref{rec-algo-nonresetting-2}, and
the parameter resetting mechanism. The entire system is given by
(\ref{dfn-Dtht})-(\ref{rec-algo-resetting-3}), where $\Phi_n$
contain all state variables of the system. The exact definitions of
all quantitites are given in Appendix B.  %%%\ref{Appendix-adaptivesystem}.

 \vskip 0.2cm

\begin{boxedminipage}{.95\textwidth}
\begin{center}
{\underline{Adaptive system}}
\begin{eqnarray}
&& D_\tht =\{ \tht: g( \tht ) \le  1, \tht_C \in D_C, \tht_F \in D_F \},  \label{dfn-Dtht} \\
&& D_R = \{ P \in \RR^{p_\tht \times p_\tht}: \kp_1 I_{p_\tht} \le P  \le \kp_2 I_{p_\tht}  \}, \label{dfn-DR} \\
&& \quad \tht_0 \in {\rm int} D_\tht  , \quad R_0 \in {\rm int} D_R ,  \label{dfn-initial} \\
&&  \Phi_{n+1} = A_\Phi (\tht_n) \Phi_n + B_\Phi (\tht_n) \eta_n ,    \label{dfn-Phi} \\
\label{rec-algo-resetting-1}
&& \tht_{n+1-} = \tht_n - \frac{1}{n+1} R_n^{-1} \e_{\tht, n+1} \e_{n+1},   \\
\label{rec-algo-resetting-2}
&& R_{n+1-} = R_n + \frac{1}{n+1} ( \e_{\tht, n+1} \e_{\tht, n+1}^T - R_n )  ,   \\
\label{rec-algo-resetting-3}
&& ( \tht_{n+1}, R_{n+1}) = \left\{ \begin{array}{cc}  ( \tht_{n+1-}, R_{n+1-}), & \tht_{n+1-} \in  D_\tht , \, R_{n+1-} \in  D_R  \\
                                        ( \tht_{0}, R_{0}),  &  {\rm otherwise.}\end{array} \right.
\end{eqnarray}
\end{center}
\end{boxedminipage}

\section{Convergence and accuracy analysis} \label{sec-convergence}
In this section, we consider the convergence of the recursive
estimation algorithm
(\ref{dfn-Dtht})-(\ref{rec-algo-resetting-3}). It is
well known that the algorithm
(\ref{dfn-Dtht})-(\ref{rec-algo-resetting-3}) can be
viewed as finite-difference equations, which has a natural
connection with ordinary differential equations (ODEs) (see
\cite{ljung77b},  \cite{ljung83}, \cite{kushner10} and
\cite{kushner03}). The ODE associated with the algorithm is given as
follows (see, e.g.,  \cite{ljung77b}, \cite{gerencser92} and
\cite{gerencser06})
\begin{subequations}  \label{dfn-ode}
\begin{align}
\frac{ \dd }{ \dd t} \tht_t & =  - R_t^{-1}  W_\tht (\tht_t)  \label{dfn-ode-tht} \\
 \frac{ \dd }{ \dd t } R_t & =   G (\tht_t) - R_t \label{dfn-ode-R}
\end{align}
\end{subequations}
for $t \ge 0$ with initial condition $( \tht_0, R_0)$, where $ W_\tht (\tht_t)$ and $G (\tht_t) $ are defined by (\ref{gradient}) and (\ref{GHessian}), respectively.

%Assume the following condition (see \cite{gerencser92,gerencser06}).
%\begin{assumption}   \label{assumption-D0}
%$\tht_0 \in D_0$, where $D_0 \subset {\rm int} D_\tht $ is a compact set such that
%\begin{equation}  \label{dfn-D0}
%\{ \tht_t: t \ge t_0, \tht_{t_0} \in D_0 \} \subset {\rm int} D_\tht .
%\end{equation}
%\end{assumption}

Assume the following condition (see  Condition 3.4 in \cite{gerencser06} and Condition C.4  in Appendix \ref{results_in_literature}). % and Remakr \ref{remarkD0}
\begin{assumption}   \label{assumption-D0}
 Let $D_0 \subseteq D_\theta$ be a compact set such that $\theta^\ast \in
 {\rm int} D_0$. We assume
 the following: {\rm(i)}~There exists a compact convex set $ D_0' \subset
 D_\theta$ such that
 \begin{equation}
 \tht (t,s,\xi) \in D_\theta \quad {\it for} ~\xi \in D_0 ~~{\it and} ~
 \tht (t,s,\xi) \in D  ~~{\it for} ~\xi \in D_0'
 \end{equation}
 for all $t \ge s \ge 0$. In addition\/ $\lim_{t \rightarrow \infty}  \tht
 (t, s, \xi) = \tht^\ast$ for $\xi \in D_0',$ and
 \begin{equation}
 \left| {\partial \over \partial \xi}  \tht (t,s,\xi)\right| \le C_0
 e^{\alpha (s-t)} \label{eqn:ODE-STAB}
 \end{equation}
 with some $C_0 \ge 1, \alpha > 0$ for all $\xi \in D_0'$ and $t \ge s
 \ge 0$.
 \/ {\rm(ii)}~We have an initial estimate $\tht_0=\xi_0$ such that for all
 $t \ge s \ge 0$ we have $ \tht  (t, s, \xi_0) \in {\rm int} D_{0}$.\/
 \end{assumption}

 \begin{remark} The condition on the existence of $D_0'$ can be removed
 if $D_\theta$ itself is convex. Note that $D_\tht$ defined by (\ref{dfn-D_tht}) is convex and hence set $D_0'$ is not needed in our paper.
\end{remark}

Moreover, another condition is imposed on the generator (\ref{input-ss}) of the input signal
\begin{assumption}   \label{assumption-input-triply}
The functions $A_z ( r(\cdot))$, $B_z (r(\cdot))$, $C_z (r(\cdot))$ and $D_z (r(\cdot))$  are triply continuously differentiable with bounded partial derivatives
up to second order on $D_\tht$.
\end{assumption}

%To present the result on convergence of the algorithm (\ref{dfn-Dtht})-(\ref{rec-algo-resetting-3}),
%let us introduce the following definitions (see \cite{gerencser89} and \cite{gerencser92}).
%

According to (\ref{system-true}) and (\ref{input-ss}), $\{ \eta_n \}$
with $\eta_n = [ e_{n+1} \,\,\, s_n ]^T$ is i.i.d.
 and
$\{ \eta_n^2 \}$ is in class $M^\ast$ (see \cite{gerencser92}). It is
also noticed that $\{ \eta_n \}$  is $L$-mixing, see Definition C.1, with respect to the
$\sgm$-algebras $(\ff_n , \ff_n^+)$ (see Appendix C). We establish the
following theorem on convergence by applying the main results in
%\cite{huang11} and
\cite{gerencser92} (see also \cite{gerencser06}),
which are listed in Appendix C. First we need to introduce the concept
of M-boundedness.

{\it Definition 4.1:} A random process $\{\bs_n \}_{n \ge 0}$  is said to be M-bounded, which is
denoted by $\bs_n = O_M (1)$, if $ M_k (\bs) = \sup_{n \ge 0} \EE^{1/k}
\big[ |\bs_n|^k \big] < \infty $ for all $1 \le k < \infty$.

Suppose that $\{ t_n \}$ is a sequence of positive numbers. We
write $\bs_n = O_M (t_n)$ if $\bs_n / t_n = O_M (1)$.

%According to Theorem 3.1 with $M=1$ \cite{huang11} and Theorem 4.1 \cite{gerencser92} (see also Theorem 3.3 \cite{gerencser06}), we have
%the following result
\begin{theorem}  \label{thm-convergence}
  % If  Assumptions \ref{assumption-stable}-\ref{assumption-D0} hold,  then  $\{ (\tht_n, R_n ) \}$ computed by the recursive algorithm (\ref{dfn-Dtht})-(\ref{rec-algo-resetting-3}) converges to $(\thtast, R^\ast)$ $a.s.$ as $n \to \infty$, where  $R^\ast = G (\thtast) $ is defined by (\ref{dfn-Rast}). Moreover, if  Assumption \ref{assumption-input-triply} also holds, then  $\{ (\tht_n, R_n ) \}$ satisfies

If  Assumptions \ref{assumption-stable}-\ref{assumption-input-triply} hold,  then  $\{ (\tht_n, R_n ) \}$ computed by the recursive algorithm (\ref{dfn-Dtht})-(\ref{rec-algo-resetting-3}) satisfies
\begin{equation}  \label{thm-convergence-OM}
\tht_n - \thtast = O_M \big( n^{-1/2} \big) \quad {\rm and} \quad R_n - R^\ast =O_M \big( n^{-1/2} \big).
\end{equation}

In particular, $\{ (\tht_n, R_n ) \}$ converges to
$(\thtast, R^\ast)$ a.s. as $n \to \infty$, where  $R^\ast = G (\thtast) $ is defined by (\ref{dfn-Rast}).
\end{theorem}

{ \it Proof:}  See Appendix D. %%%\ref{proof_of_convergence}.

Let the input signal denoted by $\{ u^\ast_n \}$ be generated by  (\ref{input-ss}) with $\tht_n = \thtast$ for all $n$.
Note that $H(n, \thtast) = - (R^\ast)^{-1} \e_{\tht, n} (\thtast)
e_{n}$ is asymptotically a wide-sense stationary process with zero mean and hence
 \cite[Condition 6.1]{gerencser06} is satisfied. Under Assumptions \ref{assumption-stable}-\ref{assumption-input-triply} (see Theorem \ref{thm-convergence}),  \cite[Theorem 6.2]{gerencser06} implies that
$ %\begin{equation}  \label{dfn-S-ast}
S^\ast = \lim_{n \to \infty} n \, \EE \left[ (\tht_n - \thtast ) (\tht_n - \thtast )^T \right]
$ %\end{equation}
exists and it satisfies the Lyapunov equation
%\begin{equation}  \label{Lyapunov-function}
$$
( A^\ast + I_{p_\tht} /2 ) S^\ast + S^\ast ( A^\ast + I_{p_\tht} /2 )^T  + P^\ast = 0
$$ %\end{equation}
with $
 A^\ast = \frac{\partial  }{\partial \tht} \big[- R^{-1} (\tht) W_\tht (\tht) \big] \Big|_{\tht = \thtast}
 = -I_{p_\tht}$ (see, e.g., \cite[(12), p.175]{lutkepohl96}) and therefore $S^\ast = P^\ast$, where
\begin{equation}  \label{dfn-P-ast}
P^\ast = \sum_{n = - \infty}^{\infty} \EE \left[H(n, \thtast) H^T(0, \thtast) \right]=\sgmaste^2 (R^\ast)^{-1}
\end{equation}
is the  covariance matrix of $\sqrt{n} (\tht_n - \thtast)$ as $n \to \infty$ when the input signal is generated by  (\ref{input-ss}) with $\tht_n = \thtast$ for all $n$.

Denoting by $\theta_n^*$ the estimate given by the off-line prediction error method under the same input excitation, it holds that $\lim_{n \to \infty} n \, \EE \left[ (\tht_n^\ast - \thtast ) (\tht_n^\ast - \thtast )^T \right]$ exists and equals \eqref{dfn-P-ast} as well \cite{ljung99}. Furthermore, the asymptotic distribution of $\sqrt{n} (\tht_n^\ast - \thtast)$ is $\NN ( 0_{p_\tht}, P^\ast)$.

Now we turn to the case where the input generator (\ref{input-ss}) is used instead of a stationary input.

\begin{theorem}  \label{thm-asy-normality}
Suppose that Assumptions \ref{assumption-stable}-\ref{assumption-input-triply} hold.
Then  $\{ (\tht_n, R_n ) \}$ computed by the recursive algorithm (\ref{dfn-Dtht})-(\ref{rec-algo-resetting-3}) satisfies
\begin{equation}  \label{asy-normal-adaptive}
\sqrt{n} \, ( \tht_n - \thtast ) \,\,\, \xrightarrow{\LL} \,\,\, \NN ( 0_{p_\tht}, P^\ast) \quad {\rm as} \quad n \to \infty,
\end{equation}
 where $P^\ast $  is the covariance matrix given by (\ref{dfn-P-ast}).
\end{theorem}

{ \it Proof:}  See Appendix E. %%%\ref{proof_of_normality}.

Comparing with the discussion before the theorem, Theorem \ref{thm-asy-normality} implies that the asymptotic distribution of $\sqrt{n} \, ( \tht_n - \thtast )$  is the same as if the input $\{ u^\ast_n \}$ is generated by  (\ref{input-ss}) with $\tht_n = \thtast$ for all $n$, and the off-line prediction error method is used.

It is observed that Assumptions \ref{assumption-stable}-\ref{assumption-noise-e} are descriptions of the nature, i.e., the LTI system (\ref{system-true}) with $\theta=\theta^\ast$, while,
in practice, Assumptions \ref{assumption-input-signal} and \ref{assumption-input-triply}
should be ensured by the input generator (\ref{input-ss}) that is
designed by the user,  which will be illustrated with an application
example in Section \ref{sec-numerical}.

%%%%
\section{Adaptive input design} \label{sec-AID}
We will now apply the results presented in the previous section and the certianty equivalance principle in \cite{gerencser05}
to the case where the input generator (\ref{input-ss}) corresponds to
the solution of an optimal input design problem. We will tailor our
results to the general frameworks in
\cite{jansson05,Geversetal:06a,Hjalmarsson:09}, consisting of the two
steps: (i) Design of the input autocorrelation sequence by way of a
semidefinite program (SDP), and (ii) spectral factorization of the
corresponding spectrum, yielding the input generator.

The main objective is to establish conditions under which Theorem
\ref{thm-asy-normality} holds, as this will then establish that
adaptive input design asymptotically achieves the same accuracy as
optimal input design in the sense that the asymptotic distribution of
$\sqrt{n}(\tht_n-\tht^*)$ for the adaptive scheme is the same as
for the off-line case using the optimal input. The assumptions related to the
input generator are Assumptions \ref{assumption-input-signal} and
\ref{assumption-input-triply}. This means that our main tasks are to establish stability of the time-varying linear system
\eqref{input-ss} and that the map from the model parameters to the
state space matrices in \eqref{input-ss} is sufficiently smooth.

For these considerations, the essential characteristics of the optimal
input design problems in the aforementioned references are that they can
be formulated as
\begin{eqnarray}  \label{dfn-min-J}
&& \min_{\hr\in\RR^{p_r}, \, \gamma\in\RR^{p_\gamma}} \gamma_1 \\
\label{dfn-min-M}
&& \text{s.t.} \,\,\, M( \hr, \gamma, \tht) \ge 0
\end{eqnarray}
The decision variable $\hr=\begin{bmatrix}\hr_1 & \ldots & \hr_{p_r}\end{bmatrix}^T\in\RR^{p_r}$ contains the coefficients in a
finite expansion of the input spectrum
\begin{align}
\label{Phiu}
\Phi_u(e^{i \om},\hr)=\sum_{k=1}^{p_r}\hr_k ({\mathcal B}_k(e^{i\omega})+{\mathcal B}_k^*(e^{i\omega})),
\end{align}
where $\{{\mathcal B}_k\}_{k=1}^{p_r}$ are stable rational basis functions. A common
choice is $B_1(z)=1/2$ and ${\mathcal B}_k(z)=z^{-(k-1)}$, $k>1$, giving an input shaping filter of FIR type.

The matrix $M(r,\gamma,\theta)$ is block diagonal where each block captures, e.g., signal constraints/criteria and model quality constraints/criteria, see below. The formulation \eqref{dfn-min-J}--\eqref{dfn-min-M}
covers both the case where a model quality measure is optimized subject to
constraints on the used signals, or the opposite formulation
(known as least-costly design \cite{Geversetal:06a}).

The auxiliary variable $\gamma=\begin{bmatrix}\gamma_1 & \ldots &
\gamma_{p_\gamma}\end{bmatrix}^T\in\RR^{p_\gamma}$ is (partly) used to
incorporate a condition that ensures that $\Phi_u$, defined
in\eqref{Phiu}, is non-negative. The latter can be ensured by the
positive real lemma (see, e.g., \cite[Lemma 2.1]{jansson05}) and
corresponds to an LMI. For the case of the basis $1/2$, $z^{-1}$,
$\ldots$ (an FIR basis), it takes the form
\begin{eqnarray}  \label{dfn-K-Q}
 \BFK (Q; \{ A_u, B_u, C_u, D_u \})  % \nonumber \\ &&
= \begin{bmatrix}  Q - A_u^T Q A_u &  - A_u^T Q  B_u \\
-B_u^T Q A_u &  - B_u^T Q B_u\end{bmatrix}
+ \begin{bmatrix}  0 & C_u^T \\ C_u  & 2 D_u \end{bmatrix} \ge 0
\end{eqnarray}
where
\begin{eqnarray}  \label{dfn-u-ABCD}
A_u & = &\begin{bmatrix} 0_{p_r-2}  &  I_{p_r-2}  \\  0  &
  0_{p_r-2}^T \end{bmatrix}, \quad B_u = \begin{bmatrix} 0_{p_r-2}\\1\end{bmatrix},  \nonumber \\
C_u & = & C_u (\hr) =  [  \hr_{p_r}  \,\,\, \cdots \,\,\, \hr_{2} ], \quad D_u = D_u (\hr) = \frac{1}{2} \hr_1
\end{eqnarray}
The unique elements of $Q=Q^T\geq 0$ are elements of $\gamma$.
The left-hand side of \eqref{dfn-K-Q} is thus one of the blocks of $M(r,\gamma,\theta)$.

Signal constraints are in terms of constraints on signal spectra, either energy constraints or frequency-by-frequency constraints.
To illustrate the expressions involved, with the input spectrum given
by \eqref{Phiu}, the input energy for an experiment of length $N$ can, using Parseval's theorem, be expressed as
\begin{align}
\label{power}
N{\mathbb E}[u^2_t]=\frac{N}{2\pi}\int_{-\pi}^{\pi}
\Phi_u(e^{i \om},\hr)\;d\omega=\sum_{k=1}^{p_r} \beta_k\; \hr_k
\end{align}
where $\beta_k=\frac{N}{2\pi}\int_{-\pi}^{\pi}({\mathcal
  B}_k(e^{i\omega})+{\mathcal B}_k^*(e^{i\omega}))d\omega$. Similarly,
the noise free output energy of a model can be expressed as
\begin{align*}
  N{\mathbb E}\left[\left(\frac{B(q,\tht_B)}{A(q,\tht_A)F(q,\tht_F)}u_t\right)^2\right]=\sum_{k=1}^{p_r} \alpha_k(\tht)\;\hr_k
\end{align*}
where
\begin{align}
  \label{alpha}
\alpha_k(\tht)=\frac{N}{2\pi}\int_{-\pi}^{\pi}\left|\frac{B(e^{i\omega},\tht_B)}{A(e^{i\omega},\tht_A)F(e^{i\omega},\tht_F)}\right|^2({\mathcal
  B}_k(e^{i\omega})+{\mathcal B}_k^*(e^{i\omega}))d\omega.
\end{align}
The blocks of \eqref{dfn-min-M} that correspond to model quality measures are affine
functions of the information matrix. Modulo a
normalization constant, the information matrix corresponds to
$G(\tht)$ defined in
\eqref{GHessian}. Employing Parseval's formula and \eqref{Phiu}, we can write
\begin{align}
  \label{Fisher}
G(\tht)=\sum_{k=1}^{p_r} \hr_k\; G_k(\tht)+G_e(\tht)
\end{align}
where
\begin{align}
    \label{Gk}
G_k(\tht)&=\frac{1}{2\pi}\int_{-\pi}^{\pi}\Gamma(e^{i\omega},\tht)\Gamma^*(e^{i\omega},\tht)({\mathcal
  B}_k(e^{i\omega})+{\mathcal B}_k^*(e^{i\omega}))d\omega\\
    \label{Ge}
G_e(\tht)&=\frac{1}{2\pi}\int_{-\pi}^{\pi}\Gamma_e(e^{i\omega},\tht)\Gamma_e^*(e^{i\omega},\tht)d\omega
\end{align}
where $\Gamma(z,\tht)$ and $\Gamma_e(z,\tht)$ are stable rational vector-valued
functions for $\tht\in D_\tht$ ,see \cite{jansson05,ljung99}.
The term $G_e$ is due to the noise excitation. Thus the information
matrix is an affine function of $r$, and hence
the blocks of \eqref{dfn-min-M} that correspond to model quality measures are affine
functions of $r$ as well.

The expressions \eqref{alpha}, \eqref{Gk}--\eqref{Ge}, are indicative of the dependence of $M$ on $\tht$. In summary,
the optimal input design frameworks of \cite{jansson05,Geversetal:06a,Hjalmarsson:09} lead to SDPs that can be written as \eqref{dfn-min-J}--\eqref{dfn-min-M}, with
\begin{align}
\label{M2}
M(\hr, \gamma, \tht)=\sum_{k=1}^{p_r} \hr_k\; M_k(\tht)+\sum_{k=1}^{p_\gamma} \gamma_k\; M_{p_r+k}(\tht)+M_{p_r+p_\gamma+1}(\tht)
\end{align}
where
\begin{align}
  \label{Mk}
M_k(\tht)=\frac{1}{2\pi}\int_{-\pi}^{\pi}\tilde{\Gamma}_k(e^{i\omega},\tht)\bar{\Gamma}_k^*(e^{i\omega},\tht)
\end{align}
where in turn $\{\tilde{\Gamma}_k(z,\tht)\}$ and $\{\bar{\Gamma}_k(z,\tht)\}$
are vector-valued rational transfer functions in $z$, with coefficients possibly depending on
$\tht$, stable on $D_\tht$.

It is clear from \eqref{M2} that \eqref{dfn-min-J}--\eqref{dfn-min-M}
is an SDP in $\hr$ and $\gamma$. Spectral
factorization of the resulting spectrum \eqref{Phiu} yields a
stable filter which we denote $G_u(z,\hr(\theta))$. Realizing this filter in state-space form
gives the input generator \eqref{input-ss}.  The filter will share poles with the basis
functions $\{{\mathcal B}_k(q)\}_{k=1}^{p_r}$. Thus it is only the
numerator coefficients that depend on $\hr$ so it can be written
\begin{align}
\label{FF}
G_u(z,\hr)=\frac{\sum_{k=0}^{m} g_k(\hr)z^{-k}}{d(z)}
\end{align}
for some fix denominator polynomial (in $z^{-1}$) $d(z)$. We can thus realize the
filter in state-space form \eqref{input-ss} using a controllable form
\cite{Kailath:80} where $A_z$ and $B_z$ are fix matrices and where
$C_z$ and $D_z$ depend linearly on the filter coefficients $\{g_k\}$.

We now summarize the adaptive input design algorithm that we will
analyze.

\begin{algthm}   \label{algorithm-adaptive}
\begin{itemize}
\item[]
\item[1)]  {\it Parametrization.} Fix the stable rational basis functions
  $\{{\mathcal B}_k(z)\}_{k=1}^{p_r}$ in the input spectrum expansion
  \eqref{Phiu}.
\item[2)]  {\it Initial estimate.} Define $D_\tht$ and $D_R$ and set $\tht_0 \in D_0 \subset {\rm int} D_\tht$, $R_0 \in {\rm int} D_R$ and $n=0$.
\item[3)] {\it Generate input process.} Take  $\{s_n \}$ to be a sequence of independent random variables satisfying (\ref{dfn-sn}).
\item[4)] {\it Input spectrum update.} Compute the optimal solution $\hr({\tht}_n)$ to (\ref{dfn-min-J})-(\ref{dfn-min-M}).
\item[5)] {\it Input filter update.} Compute the corresponding stable minimum phase input filter \eqref{FF} ($G_u(z,\hr(\tht_n))$) by spectral factorization of the corresponding input spectrum $\Phi_u (e^{i \om}, \hr(  {\tht}_n))$.
\item[6)] {\it Input generator update.} Compute the controllable state-space realization of transfer function $G_u(z,\hr(\tht_n))$.
\item[7)] {\it Measurement update.} Compute and apply the input signal $u_{n+1}$  generated by (\ref{input-ss})  to the true system and collect a new measurement $ y_{n+1} $ from the true system.
\item[8)]  {\it Parameter estimate update.} The updated recursive estimate
\begin{align*}
\tht_{n+1} =   [ \tht_{A,n+1}^T \,\,\, \tht_{B,n+1}^T \,\,\, \tht_{F,n+1}^T\,\,\,  \tht_{C,n+1}^T \,\,\, \tht_{D,n+1}^T ]^T
\end{align*}
is computed by (\ref{dfn-Dtht})-(\ref{rec-algo-resetting-3}).
\item[9)] {\it Iterate.} Replace $n$ by $n+1$ and go to step 4).
\end{itemize}
\end{algthm}

For the above algorithm we have the following result.

\begin{theorem}   \label{thm-convergence-adaptive}
Suppose that
  \begin{itemize}
    \item[(i)] $M(r,\gamma,\theta)$ in \eqref{dfn-min-M} is given by
  \eqref{M2}--\eqref{Mk}, where $\{\tilde{\Gamma}_k(z,\tht)\}$ and $\{\bar{\Gamma}_k(z,\tht)\}$ are vector-valued rational transfer functions in $z$, with coefficients possibly depending on $\tht$, stable on $D_\tht$.

    \item[(ii)]  Problem \eqref{dfn-min-J}-\eqref{dfn-min-M} is well posed in the sense that
  for each $\theta\in D_\theta$, the solution is bounded from below. Assume also that
  \eqref{dfn-min-M}   is strictly feasible for any $\theta$ in $D_\theta$.

\item[(iii)]  Problem \eqref{dfn-min-J}-\eqref{dfn-min-M} has a unique solution for every $\theta\in D_\theta$.
\item[(iv)]  Assumptions \ref{assumption-stable}-\ref{ass:jointstab} and \ref{assumption-unique-solution}-\ref{assumption-D0} hold, and the input is generated with $\{s_n\}$ satisfying the conditions in Assumption \ref{assumption-input-signal}.
\end{itemize}
  Then $\{ \tht_n \}$ generated by
  Algorithm  \ref{algorithm-adaptive} satisfies
\begin{equation}
\tht_n - \thtast = O_M \big( n^{-1/2} \big) \quad {\rm and} \quad R_n - R^\ast =O_M \big( n^{-1/2} \big).
\end{equation}
In particular,
\begin{align*}
 \tht_n \to \thtast  \;\;\; \text{almost surely as }n \to \infty.
\end{align*}
Furthermore,
\begin{align*}
\sqrt{n } \, ( \tht_n - \thtast ) \,  \xrightarrow{\LL} \,  \NN (0_{p_\tht}, P^\ast)
\end{align*}
where $P^\ast $  is the covariance matrix given by
(\ref{dfn-P-ast}), i.e. the covariance matrix obtained when an input,
$\{u_n^*\}$ say, having having the optimal input spectrum $\Phi_u(e^{i
  \om},\hr(\theta^*))$ is used.

Finally, with $\{z_n\}$ and $\{\tilde{z}_{n}\}$ denoting the input
signal $\{u_n\}$ or a stably filtered version of the input (such as,
e.g., the output $\{y_n (\tht ) \}$) when  Algorithm  \ref{algorithm-adaptive}
is operating, it holds that the limit of
\begin{align}
  \label{samplecov}
\frac{1}{n}\sum_{k=1}^{n} z_k\tilde{z}_{k-\tau}^2
\end{align}
exists almost surely for any integer $\tau$. The limit equals the
corresponding correlation for the same signals when an optimal input
$\{u_n^*\}$ is used throughout the entire experiment.
\end{theorem}
{\it Proof:} See Appendix F.

\begin{remark}
  By the arguments after Theorem \ref{thm-asy-normality}, it follows that
Theorem  \ref{thm-convergence-adaptive} shows that the adaptive Algorithm \ref{algorithm-adaptive} asymptotically recovers the same accuracy as using the optimal input during the experiment together with the off-line prediction error method.
\end{remark}

\begin{remark}
It follows from \eqref{samplecov} that the sample input power
\begin{align}
\label{sip}
\bar{u}_n^2:=\frac{1}{n}\sum_{k=1}^{n} u_k^2
\end{align}
converges almost surely to the power of the optimal input signal.
\end{remark}
\begin{remark}
The condition on well-posedness is not restrictive. For example, it is
trivially satisfied for the common objective of minimizing some
measure of the experimental effort, e.g. the input energy.
\end{remark}

\begin{remark}
  Assumption \ref{assumption-unique-solution} implies that the solution to
  (\ref{dfn-min-J})-(\ref{dfn-min-M}) has to correspond to a non-zero input spectrum for any $\theta$ in $D_\theta$.
\end{remark}

Theorem \ref{thm-convergence-adaptive} requires strict
feasibility of the SDP
\eqref{dfn-min-J}--\eqref{dfn-min-M}. In the next lemma we
establish that this holds generally for the constraints used in
\cite{jansson05,Geversetal:06a,Hjalmarsson:09}. We state the results
for the commonly used FIR basis, but the results are
straightforward to extend to a general stable rational set of basis functions.

\begin{lemma}
  \label{lem:feas}
Let $Z$ be a positive (semi-)definite matrix.
Then the LMI \eqref{dfn-K-Q} associated with the positivity condition \eqref{dfn-u-spectrum-positivity}
and the quality constraint
\begin{align}
  \label{qualcon}
G(\theta)\geq Z
\end{align}
are strictly feasible.
\end{lemma}
    {\it Proof:} See Appendix G.

\begin{remark}
Not all quality constraints in
\cite{jansson05,Geversetal:06a,Hjalmarsson:09} are of the type
\eqref{qualcon}. For example, \cite{jansson05} employ quality
constraints of the type
\begin{align}
  \label{jan}
  \begin{split}
  \mu-{\rm Tr}Z&\geq 0\\
  \begin{bmatrix}
    Z & V^* \\
    V & G
  \end{bmatrix}
  &\geq 0
  \end{split}
\end{align}
where $\mu$ and $V$ are fix quantities, and where $Z=Z^T\in\RR^{p_z\times p_z}$ is an auxiliary
variable. If we take $Z=\mu/(2p_z)I$, Schur complement give that
\begin{align}
  \label{Gschur}
G(\theta)-VZ^{-1}V^*=G(\theta)-\frac{2p_z}{\mu}\; VV^*>0
\end{align}
implies strict inequalities in \eqref{jan}, i.e. strict
feasability. The condition \eqref{Gschur} is of the type
\eqref{qualcon} and hence Lemma \ref{lem:feas} applies also to
\eqref{jan}.
\end{remark}

\begin{remark}
It is straightforward to extend Theorem \ref{thm-convergence-adaptive}
to the case where the system operates in closed loop with a fix
stabilizing LTI controller, and the experiment design problem concerns
designing the optimal reference signal. The expressions for signal
spectra and the information matrix become more involved, but retain
the structure \eqref{M2}--\eqref{Mk} that we rely on for the theorem.
\end{remark}

\begin{remark}
For input design problems where some signal size measure is the
objective function, the first phase of Algorithm
\ref{algorithm-adaptive} may generate excessive excitation
if the trajectory of the parameter estimate $\{\tht_n\}$ passes
through models that correspond to systems that are difficult to
identify, i.e. require large signal sizes in order to achieve the
quality specified by \eqref{dfn-min-J}--\eqref{dfn-min-M}. A practical
way to avoid this is to limit the signal size in an initial phase.
%As mentioned already, one standard class of optimal input design
%problems corresponds to minimizing some signal measure. For practical
%reasons it may be desirable to also in this case incorporate a
%constraint on the 'signal size', for example the allowed input power.
%This may result in that the optimization problem
%\eqref{dfn-min-J}-\eqref{dfn-min-M}
%may be infeasible for certain $\theta$. However, it is
%straightforward to modify Algorithm \ref{algorithm-adaptive} so that
%in such a situation a fix input filter (e.g. $F(q,r)=1$) is used.
%It can be shown that Theorem \ref{thm-convergence} will then still
%apply so that $\{\theta_n\}$ will approach $\theta^*$.
%Now, if there is a unique solution to the optimization problem at
%$\theta^*$, by continuity there will be a unique solution in a
%neighborhood of $\theta^*$. But $\{\theta_n\}$ eventually
%has to enter this neighborhood (and remain there thereafter almost surely), and% then
%Theorem \ref{thm-convergence-adaptive} will apply.
%Thus such a modification of the algorithm will not alter the conclusions of
%Theorem \ref{thm-convergence-adaptive}. It is the authors experience
%that such a bound can be useful in an initial phase of the experiment
%when the estimate exhibits large fluctuations due to poor
%information. This will be used in the example in the next section.
\end{remark}

%%%%%%%%%%%%%%%%%%%%%%%%%%%%%%%%%%%%%%%%%%%%%%%%%%%%%%%%%%%%%%%%%%%%%%%%%%%%%%%%
\section{Numerical illustration: $\LL_2$-gain estimation} \label{sec-numerical}
 The problem of $\LL_2$-gain estimation for FIR systems has been
 studied in  \cite[Section 6]{gerencser09}. As illustration of
 Algorithm \ref{algorithm-adaptive}, we extend this study to two cases
 where the dynamics still is of finite impulse response type,
 i.e. $A^\ast=F^\ast=1$, but where a noise model is required. The first
 case has true noise polynomials $C^\ast=1$ but $D^\ast\neq 1$
 and corresponds to a special case of an ARARX system. For an ARARX model
 structure, Assumption \ref{ass:jointstab} is trivially satisfied, see
 Remark \ref{Remark-ARARX}. As shown below, we can also impose
 a condition ensuring Assumption \ref{assumption-unique-solution}.
 The second case has $C^\ast\neq 1$ but $D^\ast=1$, i.e. a MAX (Moving
 Average with eXogenous input) system. For this case we cannot a priori
 guarantee Assumption \ref{assumption-unique-solution}, and we will
 also show that the conditions in Assumption  \ref{ass:jointstab} can
 be relaxed without affecting the performance.

 \subsection{$\LL_2$-gain estimation of ARARX systems}
 In this section, we
 consider a class of ARARX systems satisfying Assumptions
 \ref{assumption-stable}-\ref{assumption-noise-e}  and with $p_a=0$,
 $p_b \ge 2$ and $ p_d \ge 1$ (see Remark \ref{Remark-ARARX}).
 As in \cite{gerencser09}, the objective is to obtain a certain
 accuracy of an estimate of the squared $\LL_2$-gain
 $$
 \| G^\ast \|_2^2 := \frac{1}{2 \pi} \int_{- \pi}^{\pi} | G^\ast (e^{i \om})|^2 \dd \om = \thtastB^T \thtastB
 $$
 of the system transfer function $G^\ast (q) = B^\ast (q) $ at the end of an experiment of length $N$, and at the same time use
 as little input power as possible. This problem can be formulated as follows (see \cite{gerencser09})
 \begin{equation}
 \begin{array}{l} \min_u \EE [ u_n^2 ]    \\
 {\rm s.t. } \; {\rm Var}[ \|\bar{G}_N\|_2^2 ] \le \gamma ,
 \end{array}    \qquad \qquad \qquad  \qquad \qquad \label{Numerical-optimal}
 \end{equation}
 where ${\rm Var}[\cdot]$ is the variance operator with respect to the underlying probability measure, $\bar{G}_N (q) = G (q,  {\tht}_{B,N})$ represents the estimated transfer function with the truncated  estimate of $\thtastB$ and the input
 signal is generated by the linear time-varying system (\ref{input-ss}).

 %%% Let $D_\tht = \{ \tht:  | \tht_B | \vee | \tht_D|  \le K_\tht \}$ with $K_\tht >0$.
 As in \cite{gerencser09}, we use an FIR basis for the input and
 set the order $m=p_r-1=p_b -1>0$ of the input generator
 (\ref{input-ss}). In this case, $\tht = \begin{bmatrix} \tht_B^T & \tht_D^T \end{bmatrix}^T$ and $r_j = \EE [ u_n u_{n-(j-1)}]$,
 $j=1,\ldots,p_r$.
 Note that (\ref{error-recursion}) and (\ref{e-derivative}) give
\begin{eqnarray*}
\ole_n (\tht ) = y_n  -\tht^T  \varphi_n(\tht ) =y_n - \tht^T \begin{bmatrix} \tu_{n-1}  \\   - \tv_{n-1} (\tht_B) \end{bmatrix}    =\tu_{n-1}^T ( \thtastB - \theta_B) - \tv_{n-1}^T (\thtastB) \thtastD +  \tv_{n-1}^T (\theta_B) \theta_D + e_n,  \nonumber \\
\ole_{\tht, n} (\tht ) = -  \varphi_n(\tht ) + \Psi_n (\tht_D) =  \begin{bmatrix} - \tu_{n-1} \\ \tv_{n-1}^T (\tht_B)  \end{bmatrix}  +  \begin{bmatrix} - \sum_{k=1}^{p_d} d_k \tu_{n-1-k}^T \\ 0   \end{bmatrix}
= \begin{bmatrix}  - D (q, \tht_D)  \tu_{n-1} \\  \tv_{n-1}^T (\tht_B)  \end{bmatrix}   \label{def-e-e-derivative}  \hspace{1.8cm}
\end{eqnarray*}
 while  (\ref{dfn-v})  yields
 \begin{eqnarray} \label{dfn-tv-numerical}
 \tv_{n-1} (\tht_B) =  \begin{bmatrix} y_{n-1} - \tu_{n-1}^T \tht_B \\ \vdots \\ y_{n-p_d} - \tu_{n-p_d}^T  \tht_B \end{bmatrix} =  \begin{bmatrix} \tu_{n-2}^T  \\ \vdots \\ \tu_{n- p_d}^T  \end{bmatrix} (\thtastB - \tht_B)  - \begin{bmatrix} \tv_{n-1}^T (\thtastB) \\ \vdots \\
   \tv_{n-p_d}^T (\thtastB)\end{bmatrix} \thtastD +  \begin{bmatrix}  e_{n-1} \\ \vdots \\ e_{n- p_d} \end{bmatrix}.
 \end{eqnarray}
Notice that, since $\{ u_n  \}$ (with $u_n =  u_n (\thtast)$) and $\{ e_n \} $ are independent, $u_j$ and $v_k (\thtastB) = \frac{1}{ D (q, \thtastD)} e_k$ are independent for all $j$ and $k$.
This implies
\begin{eqnarray*}
\lefteqn{  \EE \left[ \left( - D (q, \tht_D)  \tu_{n-1} \right)  \tv_{n-1}^T (\theta_B) \theta_D \right] =\EE \Big[ \left( - D (q, \tht_D)  \tu_{n-1} \right)  \sum_{k=1}^{p_d} d_k \sum_{j=1}^{p_b} u_{n-k-j} (b_j^\ast- b_j) \Big] } \nonumber \\
 && = \EE \Big[ \left( - D (q, \tht_D)  \tu_{n-1} \right)  \sum_{j=1}^{p_b} (b_j^\ast- b_j)   \sum_{k=1}^{p_d} d_k  u_{n-j-k} \Big]
=  \EE \Big[ \left( - D (q, \tht_D)  \tu_{n-1} \right)  \Big(  \sum_{k=1}^{p_d} d_k \tu_{n-j-k}^T \Big) ( \thtastB - \theta_B) \Big].
\end{eqnarray*}

In the limit $n \to \infty$, (\ref{normal-equation}) is given as
 \begin{eqnarray} \label{normal-equation-numerical}
 \lefteqn{ \EE [ \ole_{\tht, n} (\tht ) \ole_n (\tht )]
 = \EE [ (- \varphi_n (\tht )  + \Psi_n (\tht ) ) \ole_n (\tht )] } \nonumber \\
 && =  \begin{bmatrix} \EE [ \big( - D (q, \tht_D)  \tu_{n-1} \big)   \tu_{n-1}^T ( \thtastB - \theta_B) ]  +    \EE [ \big( - D (q, \tht_D)  \tu_{n-1}  \big)  \tv_{n-1}^T (\theta_B) \theta_D ]  \\
       \EE [ \tv_{n-1} (\tht_B) \tu_{n-1}^T(\tht ) ] (\thtastB - \tht_B) + \EE [ \tv_{n-1} (\tht_B) \tv_{n-1}^T (\tht_B)] \tht_D
       -  \EE [ \tv_{n-1} (\tht_B) \tv_{n-1}^T (\thtastB)] \thtastD
       \end{bmatrix}  \nonumber \\
 && = 0,
 \end{eqnarray}
 that is,
 \begin{subequations}  \label{neq-1}
 \begin{align}
 & \EE \Big[ \big( D (q, \tht_D)  \tu_{n-1}  \big)  \big(D (q, \tht_D)  \tu_{n-1}^T  \big)    \Big]   ( \thtastB - \theta_B) =0 , \label{neq-1a} \\
 & R_{B}^T (\tht_B - \thtastB) + \EE [ \tv_{n-1} (\tht_B) \tv_{n-1}^T (\tht_B)] \tht_D
       -  \EE [ \tv_{n-1} (\tht_B) \tv_{n-1}^T (\thtastB)] \thtastD =0, \label{neq-1e}
 \end{align}
 \end{subequations}
where
 \begin{equation} \label{dfn-RBD}
 R_{B}    %%%%=R_{BD} (\tht_B, r)
 = \begin{bmatrix}
 \sum_{j=1}^{p_b} r_{j+1} (b_j - b_j^\ast)  & \sum_{j=1}^{p_b} r_{j+2} (b_j - b_j^\ast)  & \cdots & \sum_{j=1}^{p_b} r_{j + p_d } (b_j - b_j^\ast) \\
 \sum_{j=1}^{p_b} r_{j+1} (b_j - b_j^\ast)  & \sum_{j=1}^{p_b} r_{j+1} (b_j - b_j^\ast)  & \cdots & \sum_{j=1}^{p_b} r_{j + p_d -1} (b_j - b_j^\ast) \\
 \vdots & \vdots & \ddots & \vdots \\
 \sum_{j=1}^{p_b} r_{j-p_b+1} (b_j - b_j^\ast) & \sum_{j=1}^{p_b} r_{j-p_b+3} (b_j - b_j^\ast) & \cdots & \sum_{j=1}^{p_b} r_{j-p_b+p_d+1 } (b_j - b_j^\ast)
 \end{bmatrix}. %% \in \RR^{p_b \times p_d}.
 \end{equation}
 It is observed that, since the input signal $\{ u_n \}$ is persistently exciting,  (\ref{neq-1a}) has the unique solution $\tht_B= \thtastB$, i.e., $b_j = b_j^\ast$ for $1 \le j \le p_b$.
 Then, in this case, (\ref{neq-1e}) gives
 \begin{equation}   \label{dfn-Rv-ast}
   R_v^\ast (\tht_D- \thtastD )=0,
 \end{equation}
 which has the unique solution $\tht_D = \thtastD $ since $ R_v^\ast=\EE [ \tv_{n-1} (\thtastB) \tv_{n-1}^T (\thtastB)] >0$ (see (\ref{P-ast-numerical}) below).
 Therefore, $\tht = \thtast$ is the unique solution to the normal equations
 (\ref{normal-equation-numerical}) on any compact set $D_\tht \subset
 \RR^{p_\tht}$ with $\thtast \in {\rm int} D_\tht$.

 \subsection{The optimization problem}  \label{subsec-optimization}

 In the identification procedure, the new input of each step is determined by the solution of the optimization problem (\ref{Numerical-optimal}).
 Obviously, with the parametrization described above, the objective function $\EE [ u_n^2 ] $ in (\ref{Numerical-optimal}) equals $\hr_1$. As suggested in \cite{gerencser09}, the variance constraint ${\rm Var}[ \|\bar{G}_N\|_2^2] \le \gamma$ may be replaced using a linear approximation of $\| G (q, \overline{\tht}_N) \|_2^2$ around the true value
 \begin{equation}  \label{numerical-GN-app}
 \| \bar{G}_N \|_2^2 = \| G^\ast \|_2^2 + 2 \thtastB^T ( {\tht}_{B,N} - \thtastB ) ( 1 + \be_N )
 \end{equation}
 where $\be_N = o(1)$ is a bounded error term such that all finite moments of $\be_N$ converge to $0$ when $\overline{\tht}_{N} - \thtast$ tends to
 $0$, which implies that the variance of the squared $\LL_2$-gain can be written as
 \begin{equation}  \label{numerical-GN-var}
 {\rm Var}[ \| \bar{G}_N \|_2^2]  = 4 \thtastB^T {\rm Cov} [ {\tht}_{B,N} ] \thtastB + {\rm tr} {\rm Cov}( {\tht}_{B,N} ) \cdot o( 1 )
 \end{equation}
 with $o( 1 ) \to 0$ as $N \to \infty$, where ${\rm Cov } [\cdot]$ is the covariance operator with respect to the underlying probability measure.
 According to Theorem \ref{thm-convergence-adaptive}, the original variance constraint may be replaced by an approximation
 \begin{equation} \label{numerical-varG-app}
 4 \thtastB^T \frac{  \sgmaste^2}{N }  (R_u^\ast)^{-1} \thtastB \le \gamma,
 \end{equation}
 where $R_u^\ast \in \RR^{p_b \times p_b}$  is the principal submatrix of $R^\ast$ and therefore
 \begin{equation} \label{P-ast-numerical}
 P^\ast = \sgmaste^2 (R^\ast)^{-1}= \sgmaste^2
 \begin{bmatrix} R_u^\ast & 0 \\ 0 &  R_v^\ast \end{bmatrix}^{-1} = \sgmaste^2 \begin{bmatrix} (R_u^\ast)^{-1} & 0 \\ 0 &  (R_v^\ast)^{-1} \end{bmatrix}.
 \end{equation}
 Inequality (\ref{numerical-varG-app}), by Schur complements, can be expressed as
 \begin{equation}  \label{numerical-R-LMI}
 \begin{bmatrix} R_u^\ast  & 2 \thtastB \\
 2 \thtastB^T  &  \frac{\gamma N}{ \sgmaste^2} \end{bmatrix} \ge 0.
 \end{equation}

 In the adaptive input design context, at each step we replace the true value $\thtast$  with
 the  estimate ${\tht}_n =  [ \, {\tht}_{B, n}^T \,\,\,  {\tht}_{D, n}^T]^T$.
 Therefore, the optimization problem that is solved at time step $n$ is given by
 %\begin{eqnarray}    \label{numerical-final_form}
 %&& \min_{ r, \, Q }  \,\, r_{1}   \\
 %&& {\rm s.t.}
 %\begin{bmatrix} R_u(\tht )  & 2 {\tht}_{B,n}  \\
 %2 {\tht}_{B,n}^T  &  \frac{\gamma N}{ {\sgm}_{e, n}^2} \end{bmatrix} \ge 0, \; \;  \BFK (Q; \{ A_u, B_u, C_u(\tht ), D_u(\tht ) \}) \ge 0, \; \; Q \ge 0 , \nonumber \\
 %&& \qquad    R_u(\tht ) + \frac{1}{2} [ R_{D}^T (\tht_{D,n}) + R_{D}
 %  (\tht_{D,n}) ] \ge \beta_{D} I_{p_b},\, R_u(r)\geq \beta_r I_{p_b} \nonumber
 %\end{eqnarray}
 \begin{eqnarray}    \label{numerical-final_form}
 && \min_{ r, \, Q }  \,\, r_{1}   \\
 && {\rm s.t.}
 \begin{bmatrix} R_u(\tht )  & 2 {\tht}_{B,n}  \\
 2 {\tht}_{B,n}^T  &  \frac{\gamma N}{ {\sgm}_{e, n}^2} \end{bmatrix} \ge 0, \; \;  \BFK (Q; \{ A_u, B_u, C_u(\tht ), D_u(\tht ) \}) \ge 0,  \nonumber \\
 && \qquad   Q \ge 0 ,  \;\;  R_u(\tht ) \ge \beta_R I_{p_b},   \nonumber
%%%, \quad r_1\leq r_{\max}\nonumber
 \end{eqnarray}
 where  $R_u(\tht )$ is the symmetric Toeplitz matrix with $r$ as first
 column and $ \beta_R$ is a   small positive number
 set to ensure the persistent excitaion condition.
% We have also included an upper bound on the $r_{max}$ on the input signal power, cf. Remark 5.5.

 As in \cite{gerencser09}, the optimization is made with the MATLAB toolbox YALMIP (\cite{gahinet95} and \cite{lofberg04}) and the solver {\rm sdpt3} (\cite{toh06}). The conditions of Theorem \ref{thm-convergence-adaptive} are satisfied for the procedure described above, which implies that
 the parameter estimates will converge to the true value almost surely and the asymptotic accuracy for the adaptive design will be the same as for the optimal input.

 %%%%%%%%%%%%%%%%%%%%%%%%%%%%%%%%%%%%%%%%%%%%%%%%%%%%%%%%%%%%%%%%%%%%%%%%%

 \begin{figure*}
 \begin{minipage}{0.48\textwidth}
 \centering
 \includegraphics[width=8cm]{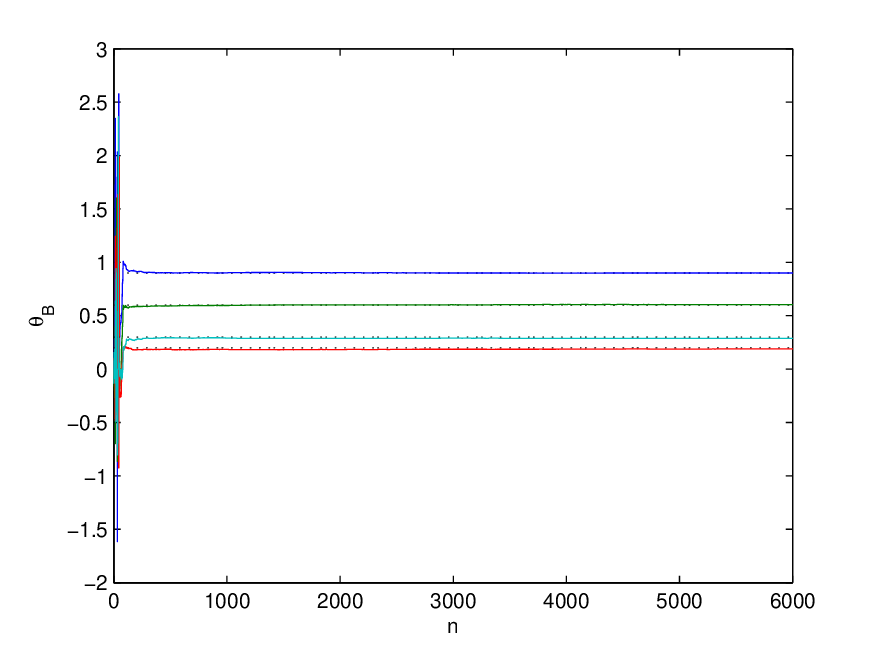}
 \vskip -0.5cm
 \caption{ {\footnotesize Solid lines: estimates of $\theta_B$ by Algorithm \ref{algorithm-adaptive}. Dotted lines: true values.} }
 \label{fig1}
 \end{minipage}%
 \hskip 0.3cm
 \begin{minipage}{0.48\textwidth}
 \centering
 \includegraphics[width=8cm]{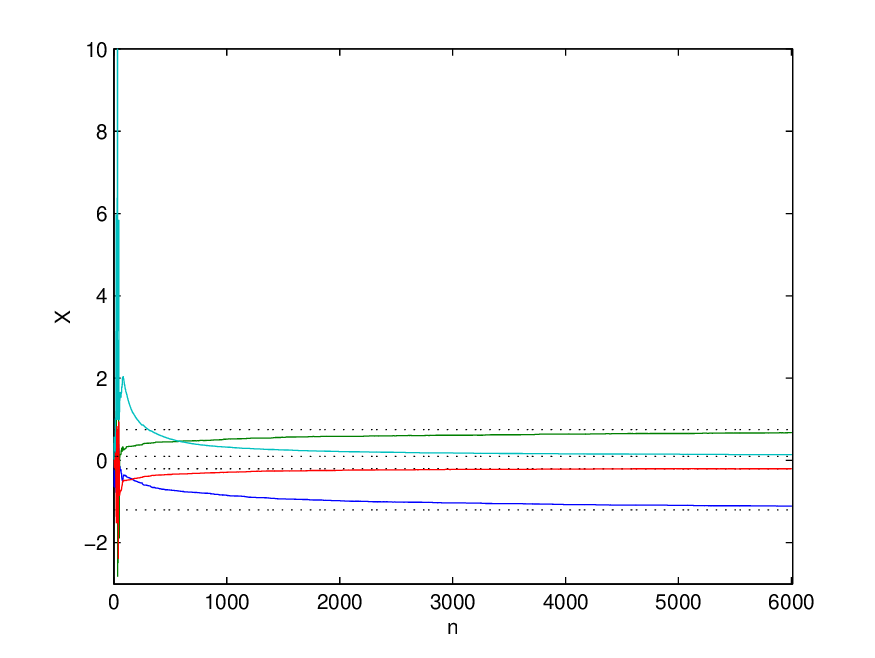}
 \vskip -0.5cm
 \caption{ {\footnotesize Solid lines: estimates of $X= [ \theta_D \,\,\, \sigma_e^2]$ by Algorithm \ref{algorithm-adaptive}. Dotted lines: true values.} }
 \label{fig2}
 \end{minipage}
 %\end{figure*}
 %
 %%%%%%%%%%%%%%%%%%%%%%%%%%%%%%%%%%%%%%%%%%%%%%%%%%%%%%%%%%%%%%%%%%%%%%%%%%
 %
 %\begin{figure*}
 \begin{minipage}{0.48\textwidth}
 \centering
 \includegraphics[width=8cm]{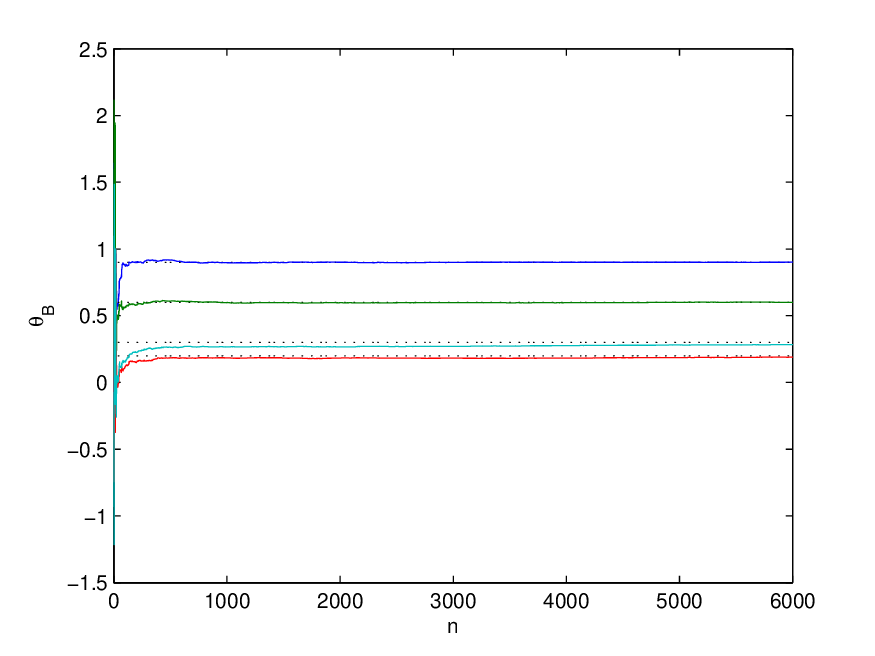}
 \vskip -0.5cm
 \caption{ {\footnotesize Solid lines: estimates of $\theta_B$ by algorithm (\ref{dfn-Dtht})-(\ref{rec-algo-resetting-3}) with  optimal input. Dotted lines: true values.} }
 \label{fig3}
 \end{minipage}%
 \hskip 0.3cm
 \begin{minipage}{0.48\textwidth}
 \centering
 \includegraphics[width=8cm]{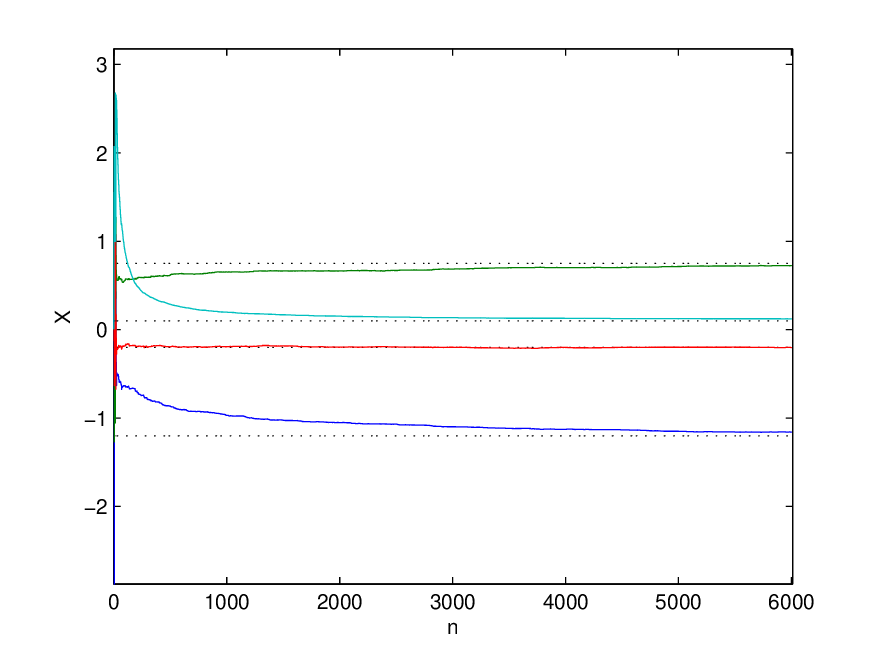}
 \vskip -0.5cm
 \caption{ {\footnotesize Solid lines: estimates of $X= [ \theta_D \,\,\, \sigma_e^2]$ by algorithm (\ref{dfn-Dtht})-(\ref{rec-algo-resetting-3}) with  optimal input. Dotted lines: true values.} }
 \label{fig4}
 \end{minipage}
 \end{figure*}
 %
 %%%%%%%%%%%%%%%%%%%%%%%%%%%%%%%%%%%%%%%%%%%%%%%%%%%%%%%%%%%%%%%%%%%%%%%%%%%%%%%%%%%
 \begin{figure*}
 \begin{minipage}{0.48\textwidth}
 \includegraphics[width=8cm]{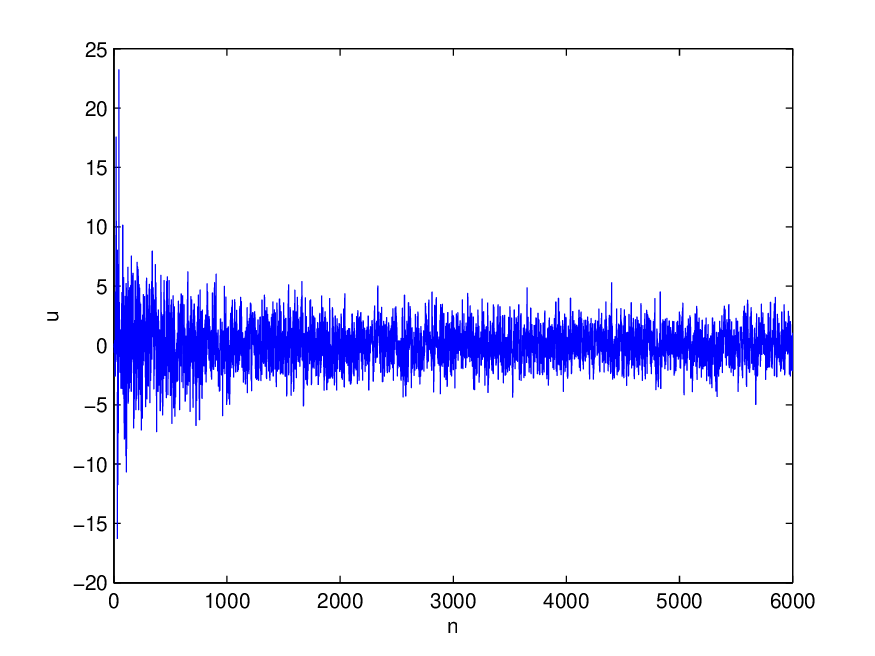}
 \vskip -0.5cm
 \caption{ {\footnotesize The realization of adaptive input signal
     $u$ for Algorithm \ref{algorithm-adaptive} corresponding to
     Figs. \ref{fig1}-\ref{fig2}.} }
 \label{fig5}
 \end{minipage}%
 \hskip 0.3cm
 \begin{minipage}{0.48\textwidth}
 \centering
 \includegraphics[width=8cm]{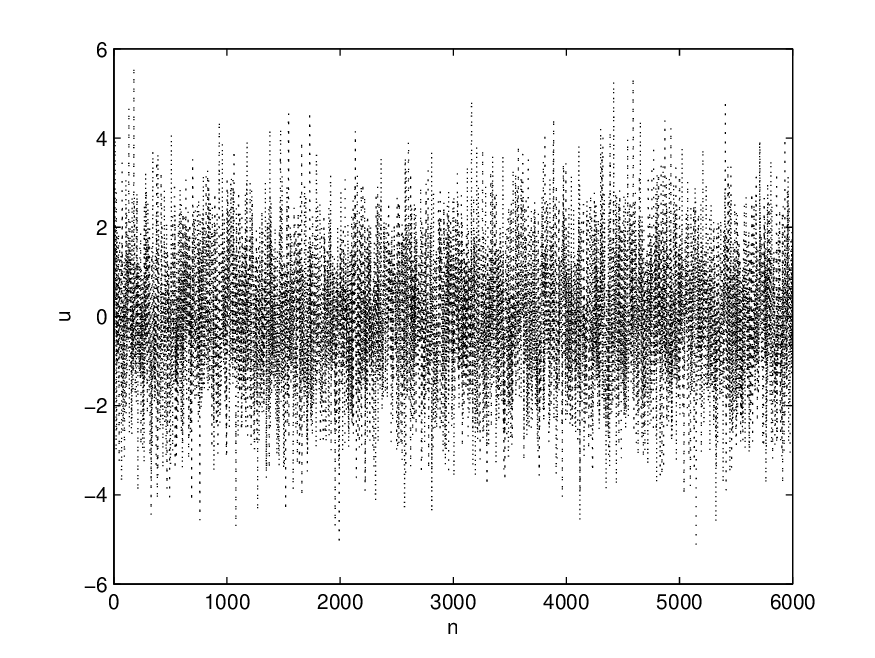}
 \vskip -0.5cm
 \caption{ {\footnotesize  The realization of optimal input signal
     $u$  corresponding to
     Figs. \ref{fig3}-\ref{fig4}.} }   \label{fig6}
 \end{minipage}
 %\end{figure*}

 \begin{minipage}{0.48\textwidth}
 \includegraphics[width=8cm]{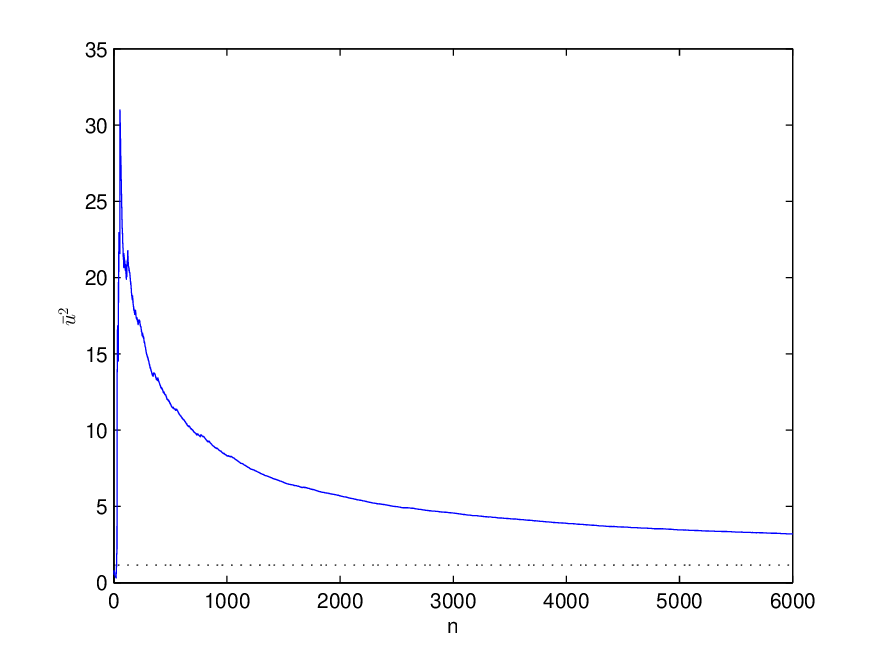}
 \vskip -0.5cm
 \caption{ {\footnotesize Solid line: the realization of sample input power
     $\bar{u}^2$ for Algorithm \ref{algorithm-adaptive} corresponding to
     Figs. \ref{fig1}, \ref{fig2} and \ref{fig5}. Dotted line: input power $r_1$ of the optimal input.} }
 \label{fig7}
 \end{minipage}%
 \hskip 0.3cm
 \begin{minipage}{0.48\textwidth}
 \centering
 \includegraphics[width=8cm]{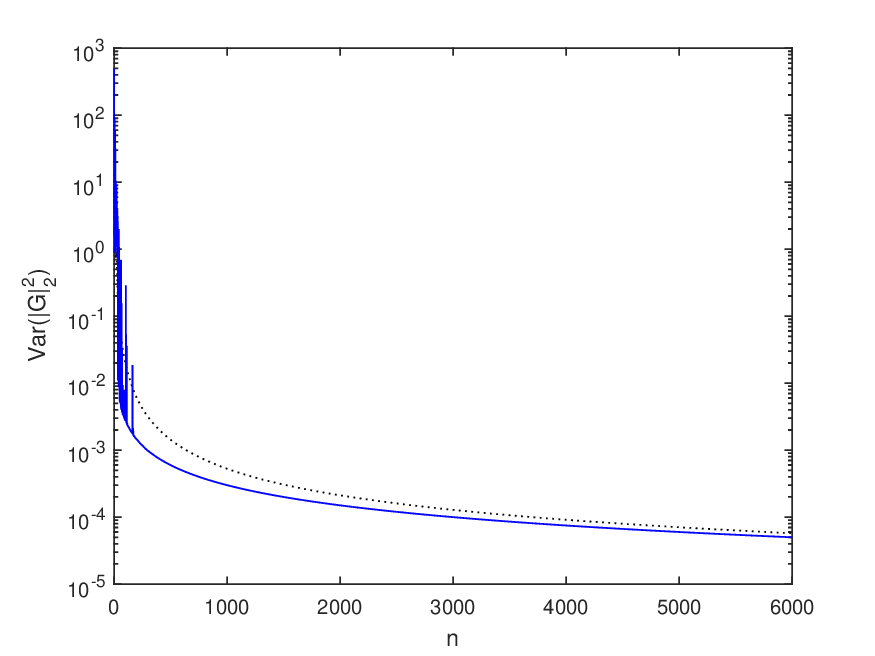}
 \vskip -0.5cm
 \caption{ {\footnotesize Variance of the estimated $\LL_2$-gain, ${\rm Var}( \|\bar{G}_N\|_2^2)$. Solid line: variance estimated from Monte Carlo simulations with the adaptive input. Dotted line: variance estimated from Monte Carlo simulations with the optimal input.} }   \label{fig8}
 \end{minipage}
 \end{figure*}

 \subsection{Simulation results}  \label{subsec-simulations}
 Generalizing the FIR numerical example in \cite{gerencser09},
 we take the true  parameters of the ARARX system  with orders $p_a=0$,
 $p_b=4$ and $p_d=3$ to be
 $\thtastB = ( 0.9 \,\,\, 0.6 \,\,\, 0.2 \,\,\, 0.3 )^T$, $\thtastD = ( -1.2 \,\,\,  0.75 \,\,\, -0.2 )^T$ and $\sgmaste^2 = 0.1$.
 As in \cite{gerencser09}, we set the order $m=p_r-1=3$ for the linear
 time-varying system (\ref{input-ss}). In the following simulations, we employ the algorithm (\ref{dfn-Dtht})-(\ref{rec-algo-resetting-3})  and choose $D_\tht = \{\tht:  | \tht_B|   \le 3, | \tht_D| \le 10 \}$ , $D_R$ with $\kp_1 =10^{-6}$ and $\kp_2 =10^{10}$, initial
 value $\tht_{0} = [ \tht_{B,0}^T \,\,\, \tht_{D, 0}^T ]^T =  0_7 $ and $R_0 = I_7$.
 %In the following simulations, we
 %employ the algorithm (\ref{dfn-Dtht})-(\ref{rec-algo-resetting-3})
 %and choose $D_\tht = \{\tht:  | \tht
 %\e (\tht)|^2  \le K_\theta \}$ with $K_\theta = 20$ , $D_R$ with
 %$\kp_1 =10^{-6}$ and $\kp_2 =10^{10}$, initial
 %value $\tht_{0} = [ \tht_{B,0}^T \,\,\, \tht_{D, 0}^T ]^T =  0_7 $ and
 %$R_0 = I_7$.

 %
 The total experiment length $N=6 \times 10^3$, the required accuracy
 $\gamma = 5 \times 10^{-5}$ and $ \beta_R = 10^{-2}$. %  and  $r_{max}=5$.
 %The total experiment length $N=4 \times 10^3$, the required accuracy
 %$\gamma = 10^{-4}$, $\beta_R = 10^{-3}$, $\hr_{max}=5$ and $\beta_D =10^{-6}$.
 % We set the input bounds  $r_{min}=0.001$ to ensure the power of input signal $r_{n,0} >0$ and $r_{max}=10^{10}$ to be soft enough that it become active throughout the simulation (cf. \cite{gerencser09}).
 Figs. \ref{fig1} and \ref{fig2} show a typical realization of
 Algorithm \ref{algorithm-adaptive} for estimates of $\theta_B$ and $
 X=\begin{bmatrix} \theta_D^T & \sigma_e^2 \end{bmatrix}^T$,
 respectively, while  Figs. \ref{fig3} and \ref{fig4} show a typical
 realization of algorithm (\ref{dfn-Dtht})-(\ref{rec-algo-resetting-3})
 with  the optimal input signal that is generated by (\ref{input-ss}) with
 parameters obtained by solving the optimization problem
 (\ref{numerical-final_form}) with $\tht_n =
 \thtast$. Figs. \ref{fig5} and \ref{fig6}
 shows the input signal $u$ for the same realizations as in
 Figs. \ref{fig1}-\ref{fig2} and Figs. \ref{fig3}-\ref{fig4}, respectively.
 The  realization of the sample input power \eqref{sip} corresponding to Figs \ref{fig1}-\ref{fig2}, as well as
 $r_1$ of the optimal input, are shown in Fig. \ref{fig7}. Fig. \ref{fig8}
 shows the variance of the estimated $\LL_2$-gain, ${\rm Var}( \|\bar{G}_N\|_2^2)$, estimated from $100$ Monte Carlo simulations with the adaptive input and the optimal input, respectively.

 \begin{figure*}
 \begin{minipage}{0.48\textwidth}
 \centering
 \includegraphics[width=7cm]{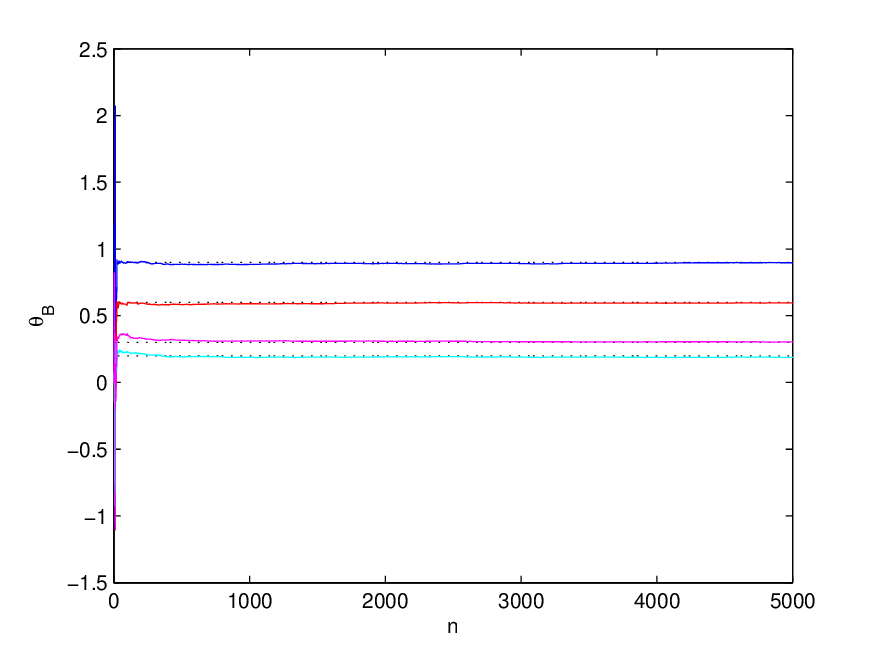}
 \vskip -0.8cm
 \caption{ {\footnotesize Solid lines:  estimates of $\theta_B$ by Algorithm \ref{algorithm-adaptive}. Dotted lines: true values.} }
 \label{fig9}
 \end{minipage}%
 \hskip 0.3cm
 \begin{minipage}{0.48\textwidth}
 \centering
 \includegraphics[width=7cm]{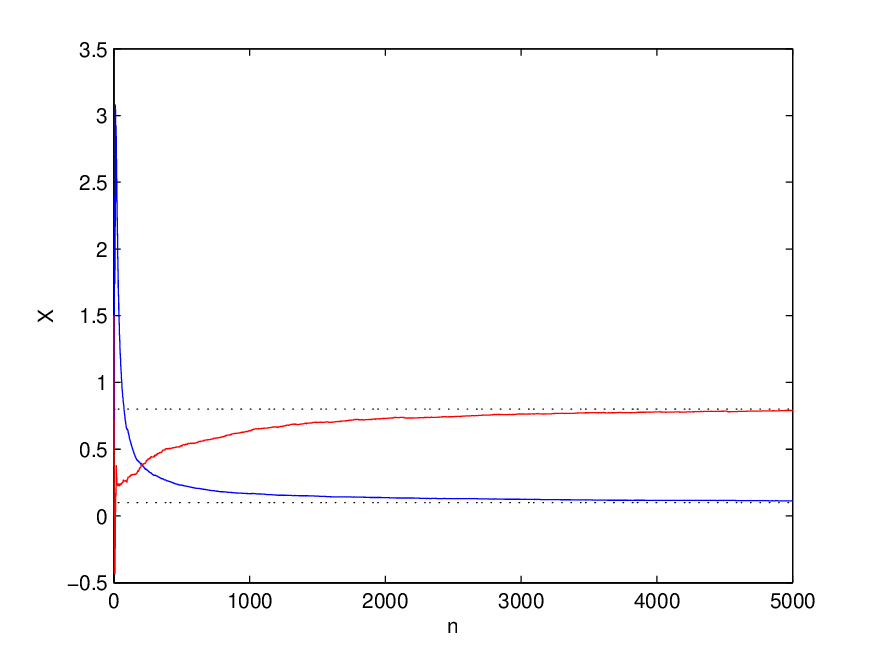}
 \vskip -0.8cm
 \caption{ {\footnotesize Solid lines:  estimates of $X=[ \theta_C \;\; \sigma_e^2 ]$ by Algorithm \ref{algorithm-adaptive}. Dotted lines: true values.} }
 \label{fig10}
 \end{minipage}
 %\end{figure*}
 \vskip 0.2cm
 \begin{minipage}{0.48\textwidth}
 \centering
 \includegraphics[width=7cm]{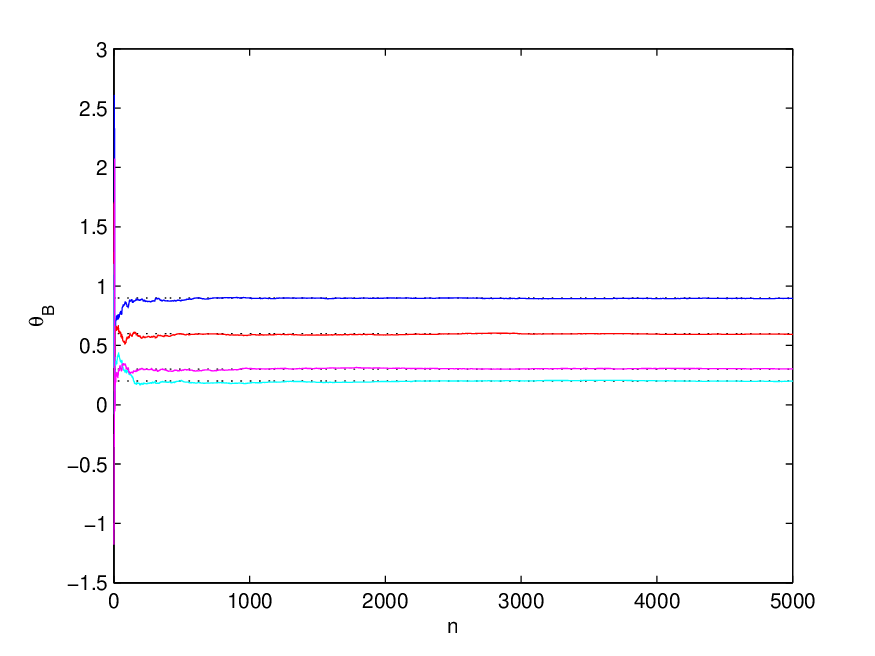}
 \vskip -0.8cm
 \caption{ {\footnotesize Solid lines: estimates of  $\theta_B$ by  algorithm (\ref{dfn-Dtht})-(\ref{rec-algo-resetting-3})  with optimal input. Dotted lines: true values.} }
 \label{fig11}
 \end{minipage}%
 \hskip 0.3cm
 \begin{minipage}{0.48\textwidth}
 \centering
 \includegraphics[width=7cm]{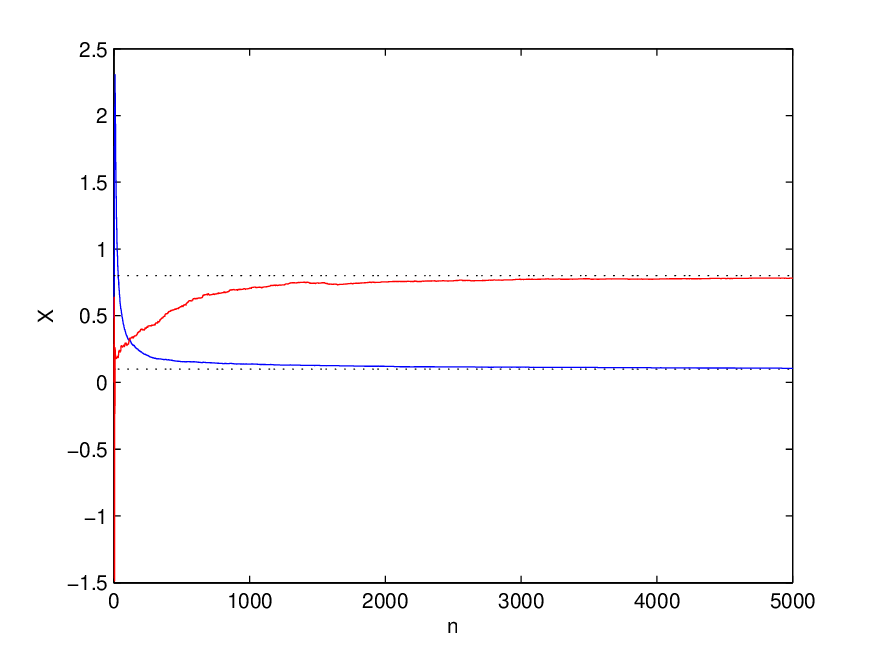}
 \vskip -0.8cm
 \caption{ {\footnotesize Solid lines: estimate of  of $X=[ \theta_C \;\; \sigma_e^2 ]$ by  algorithm (\ref{dfn-Dtht})-(\ref{rec-algo-resetting-3})  with optimal input. Dotted lines: true values.} }
 \label{fig12}
 \end{minipage}
 \end{figure*}

 \begin{figure*}
 \begin{minipage}{0.48\textwidth}
 \includegraphics[width=8cm]{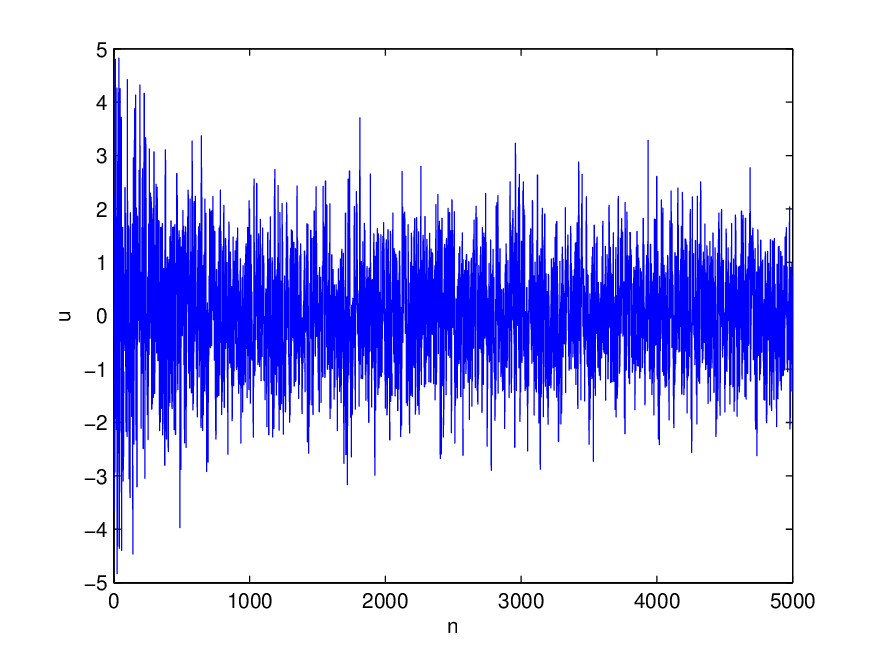}
 \vskip -0.5cm
 \caption{ {\footnotesize The realization of adaptive input signal
     $u$ for Algorithm \ref{algorithm-adaptive} corresponding to
     Figs. \ref{fig9}-\ref{fig10}.} }
 \label{fig13}
 \end{minipage}%
 \hskip 0.3cm
 \begin{minipage}{0.48\textwidth}
 \centering
 \includegraphics[width=8cm]{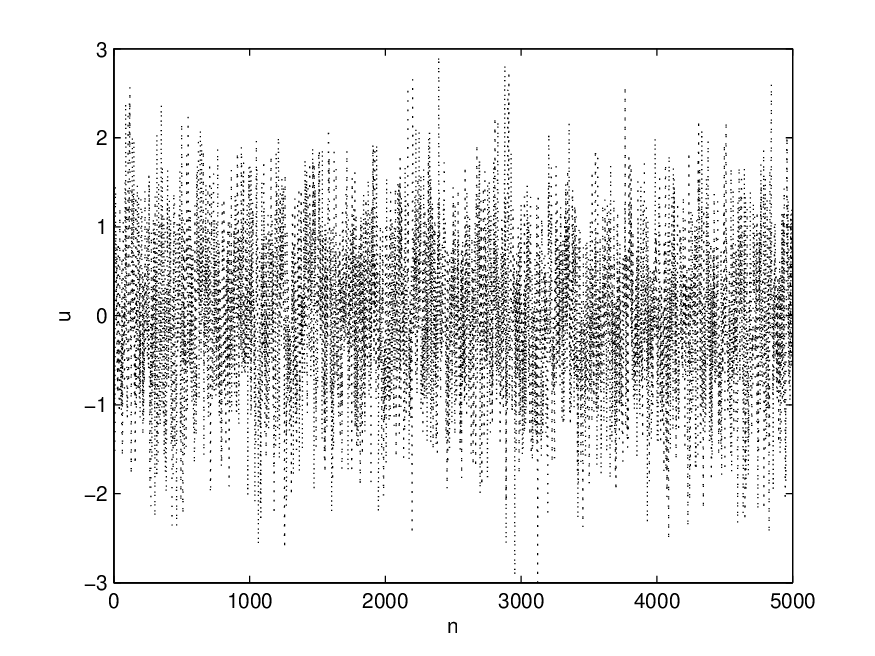}
 \vskip -0.5cm
 \caption{ {\footnotesize  The realization of optimal input signal
     $u$  corresponding to
     Figs. \ref{fig11}-\ref{fig12}.} }   \label{fig14}
 \end{minipage}
 \begin{minipage}{0.48\textwidth}
 \includegraphics[width=8cm]{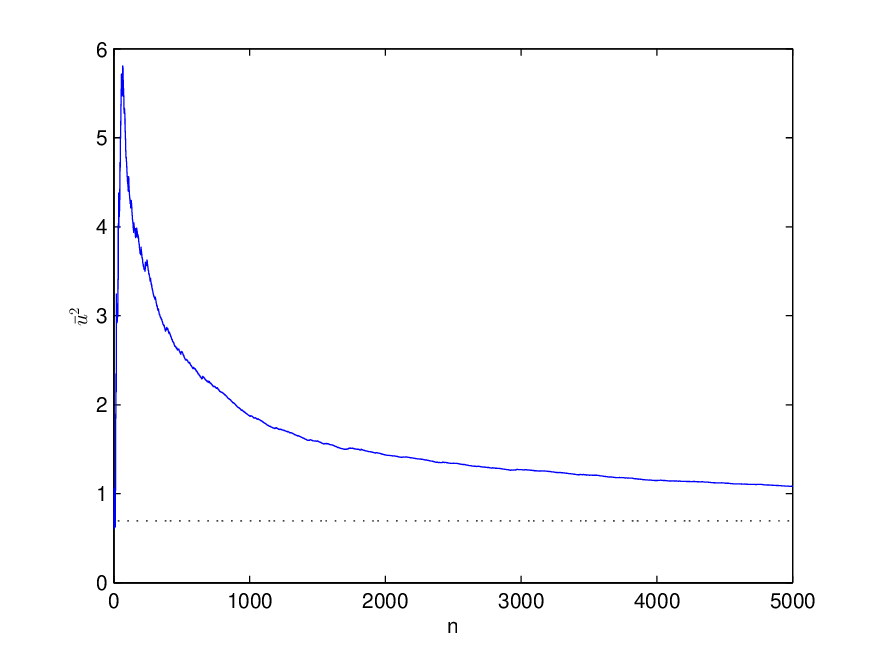}
 \vskip -0.5cm
 \caption{ {\footnotesize Solid line: the realization of sample input power  $\bar{u}^2$ for Algorithm \ref{algorithm-adaptive} corresponding to
     Figs. \ref{fig9}, \ref{fig10} and \ref{fig13}. Dotted line: input power $r_1$ of the optimal input.} }
 \label{fig15}
 \end{minipage}%
 \hskip 0.3cm
 \begin{minipage}{0.48\textwidth}
 \centering
 \includegraphics[width=8cm]{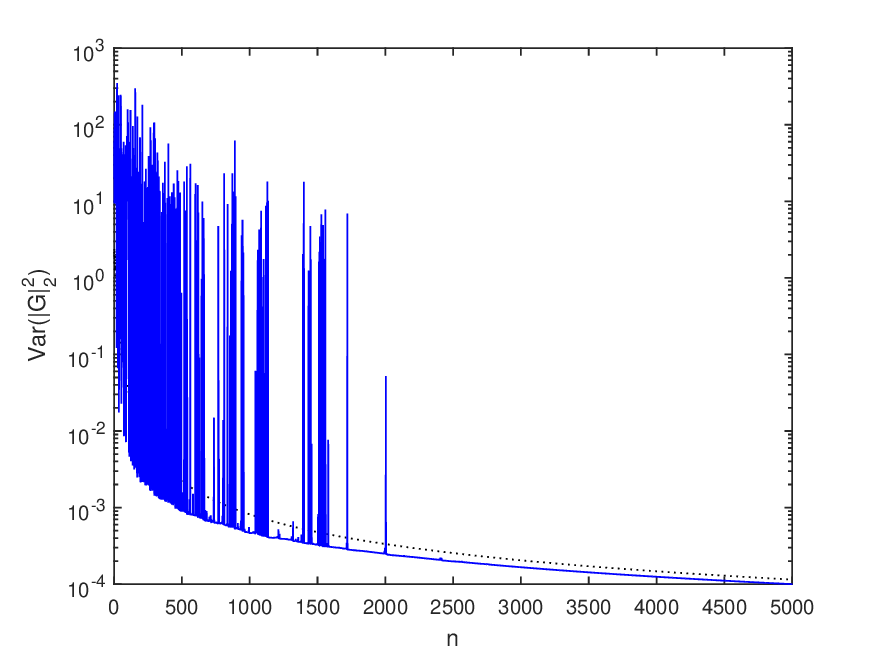}
 \vskip -0.5cm
 \caption{ {\footnotesize Variance of the estimated $\LL_2$-gain, ${\rm Var}( \|\bar{G}_N\|_2^2)$. Solid line:  variance estimated from Monte Carlo simulations with the adaptive input. Dotted line:  variance estimated  from Monte Carlo simulations with the optimal input.} }   \label{fig16}
 \end{minipage}
 \end{figure*}

 \subsection{Simulation results for a MAX system}  \label{subsec-simulations_ARMAX}
  As noted in Remark \ref{Remark-ARARX2}, the conditions imposed by
  Assumption 4  are
  restrictive.
  However, we will now illustrate that the  proposed algorithm may
  work even if the conditions in Assumption 4 are not
  satisfied. To this end, let us generalize the FIR numerical example in
  \cite{gerencser09} to a MAX system.  The true  parameters of the MAX
  system  with orders  $p_b=4$ and $p_c=1$ are
  $\thtastB = \begin{pmatrix} 0.9 & 0.6 & 0.2 & 0.3 \end{pmatrix}^T$, $\thtastC =c_1^\ast = 0.8$ and $\sgmaste^2 = 0.1$.
 % In the following simulations,   choose $D_\tht = \{\tht: |\tht_B | \le 10, |c_1| \le K_c\}$, $D_R$ with $\kp_1 =10^{-6}$ and $\kp_2 =10^{10}$, and initial
 %values $\tht_{0} = [ \tht_{B,0}^T \,\,\, \tht_{C, 0} ]^T =  ( 0 \,\,\, 0 \,\,\, 0 \,\,\, 0 \,\,\, 0 )^T \in D_0$ and $R_0 = I_5$. Moreover, $N=2000$, $\gamma = 0.0005$.
 %Note that, as required by Assumption 4, $K_C = 0.9999<1$ for $D_C = \{\theta_C:\;
 %| \theta_C| \le K_c\}$ in [39]. But, from simulations, it appears that
 %$K_c$ can be chosen larger or as an increasing sequence as $\{ M_n
 %\}_{n \ge 0}$ in \cite{chen10} and \cite{huang14}. Here, we take $K_C =5$, which is much
  %larger than what is required by Assumption 4.
 In the following simulations,  we have used $D_\tht = \{\tht: |\tht_B | \le 3, |c_1| \le K_C\}$, $D_R$ with $\kp_1 =10^{-6}$ and $\kp_2 =10^{10}$, and initial
 values $\tht_{0} = [ \tht_{B,0}^T \,\,\, \tht_{C, 0} ]^T =  ( 0 \,\,\, 0 \,\,\, 0 \,\,\, 0 \,\,\, 0 )^T \in D_0$ and $R_0 = I_5$. Moreover, $N=5 \times 10^3$ and $\gamma =10^{-4}$.
 Note that, as required by Assumption 4, $K_C = 0.9999<1$ for $D_C = \{ | \theta_C| \le K_C \}$ in \cite{huang12}. But, from simulations, it appears that $K_C$ can be chosen larger. Here we take $K_C =5$, which is much larger than what is required by Assumption 4.   Notice also that
 Assumption \ref{assumption-unique-solution} can not be ensured in this
 example. Figs. \ref{fig9} and
 \ref{fig10} originate from a typical realization of  Algorithm
 \ref{algorithm-adaptive}, while Figs. \ref{fig11} and \ref{fig12} are
 typical realizations from the same algorithm, save for that the
 optimal input is used.   The realization of sample input
 power $\bar{u}^2 $ corresponding to Figs
  \ref{fig9}-\ref{fig10}, as well as  $r_1$ of the optimal input, are
  shown in Fig. \ref{fig13}. Fig. \ref{fig14}
 shows the variance of the estimated $\LL_2$-gain, ${\rm Var}(
 \|\bar{G}_N\|_2^2)$, estimated from $100$ Monte Carlo simulations with
 the adaptive input and the optimal input, respectively. We see that
 also in this case Algorithm \ref{algorithm-adaptive} performs well,
 despite that some of the assumptions are not satisfied. Thus the
 algorithm exhibit some degree of robustness.

\section{Conclusion}
\label{sec:concl}
This paper presents sufficient conditions for consistency
of an adaptive system consisting of a SISO LTI
system, a recursive prediction error estimator and an input generator
which uses the parameter estimates. The asymptotic distribution of the
resulting parameter estimates has been derived as well.

As an application, we have proposed an adaptive input design method
for stable LTI systems based on the certainty equivalence
principle. This is a formal development of the scheme outlined in
\cite{gerencser05}, which establishes convergence and
asymptotic efficiency. The asymptotic theory is backed-up by a
finite-sample simulation study.

%It is our future work to develop the underlying theory and remove the restritive conditions (\ref{dfn-Vc})-(\ref{dfn-Vf}).

%%%%%%%%%%%%%%%%%%%%%%%%%%%%%%%%%%%%%%%%%%%%%%%%%%%%%%%%%%%%%%%%%%%%%%%%%%%%%%%%
\section*{Appendix A. Proof of Lemma \ref{G-positive-definite}  }   \label{proof_of_G_positive}
Clearly, given any $\tht \in  D_\tht$, $ \EE [ \ole_{\tht, n} (\tht ) \ole_{\tht, n}^T (\tht ) ] \in \RR^{ p_\tht  \times p_\tht }$ is a symmetric positive semidefinite matrix.
Furthermore,  $ \EE [ \ole_{\tht, n} (\tht ) \ole_{\tht, n}^T (\tht ) ] = \EE [ \varphi_n (\tht ) \varphi_n^T (\tht ) ]$ is not symmetric positive definite if and only if there exists a nonzero vector $\nu \in \RR^{p_\tht }$ such that $\nu^T \EE [ \ole_{\tht, n} (\tht ) \ole_{\tht, n}^T (\tht )] \nu =  \EE [ \nu^T \ole_{\tht, n} (\tht ) \ole_{\tht, n}^T (\tht ) \nu]  = \EE [|\nu^T \ole_{\tht, n} (\tht ) |^2 ] = \EE [|\nu^T \varphi_n (\tht )|^2 ] =0 $.

Let
$$
\bG^\ast (q) = \frac{B^\ast (q) }{ A^\ast (q) F^\ast (q)} \quad {\rm and } \quad \bH^\ast (q) = \frac{C^\ast (q) }{ A^\ast (q) D^\ast (q)},
$$
then $y_n (\tht )  = \bG^\ast (q)u_n + \bH^\ast (q) e_n$. By (\ref{e-derivative}), we observe that
\begin{equation}  \label{e-derivative-u-e}
\ole_{\tht, n} (\tht) = \ole_{\tht, n} (\tht ) =  \begin{bmatrix} F_u (q, \tht ) & F_e (q, \tht )  \end{bmatrix}  \begin{bmatrix}u_n & e_n \end{bmatrix}^T =  F_u (q, \tht)u_n + F_e (q, \tht) e_n
\end{equation}
and hence $\EE [ \ole_{\tht, n} (\tht) \ole_{\tht, n}^T(\tht) ] $ are continuous on $ D_\tht$ since both $F_u (q, \tht)$ and $F_e (q, \tht)$  are continuous on $ D_\tht$, where
\begin{eqnarray*}
&& F_u (q, \tht) = \begin{bmatrix}  F_{u \ty}(q) & -F_{u \tu} (q) & F_{u \tw} (q, \tht) & -F_{u \te} (q, \tht) & F_{u \tv} (q, \tht)  \end{bmatrix}^T, \\
&& F_e (q, \tht) = \begin{bmatrix}  F_{e \ty}(q) &   0_{p_b}^T         &   0_{p_f}^T         &- F_{e \te} (q, \tht) & F_{e \tv} (q, \tht) \end{bmatrix}^T, \\
&& F_{u \ty}(q) =  \left[ q^{-1}  \;\;\; \cdots \;\;\; q^{-p_a}  \right] \bG^\ast(q) ,  \quad F_{u \tu} (q)= \left[ q^{-1} \;\;\; \cdots \;\;\; q^{-p_b}  \right], \\
&& F_{u \tw}(q, \tht) =  \left[ q^{-1}  \;\;\; \cdots \;\;\; q^{-p_f} \right] F_{wu}(q, \tht), \;\;F_{wu}(q, \tht) = \frac{B(q, \tht_B)}{F(q, \tht_F)} , \\
&& F_{u \te}(q, \tht) =  \left[ q^{-1}  \;\;\; \cdots \;\;\; q^{-p_c}  \right] F_{\e u}(q, \tht), \;\;
     F_{\e u}(q, \tht) = \frac{D(q, \tht_D)}{C(q, \tht_C)} F_{vu}(q, \tht), \\
&& F_{u \tv}(q, \tht) =  \left[ q^{-1}  \;\;\; \cdots \;\;\; q^{-p_d} \right] F_{vu}(q, \tht), \;\;F_{vu}(q, \tht) = A(q, \tht_A) \bG^\ast (q)- F_{wu}(q, \tht), \\
&& F_{e \ty}(q) =  \left[ q^{-1}  \;\;\; \cdots \;\;\; q^{-p_a}  \right] \bH^\ast(q) ,   \\
&& F_{e \te}(q, \tht) =  \left[ q^{-1}  \;\;\; \cdots \;\;\; q^{-p_c} \right] F_{\e e}(q, \tht), \;\; F_{\e e}(q, \tht) = \frac{D(q, \tht_D) A(q, \tht_A) }{C(q, \tht_C)} \bH^\ast (q), \\
&& F_{e \tv}(q, \tht) =  \left[ q^{-1}  \;\;\; \cdots \;\;\; q^{-p_d} \right] F_{v e}(q, \tht), \;\; F_{v e}(q, \tht) =  A(q, \tht_A) \bH^\ast (q) .
\end{eqnarray*}
Particularly, we have
\begin{equation}  \label{e-derivative-u-e2}
\ole_{\tht, n} (\thtast) =  F_u (q, \thtast)u_n + F_e (q, \thtast) e_n,
\end{equation}
where
$ F_u (q, \thtast) = \begin{bmatrix}  F_{u \ty}(q) & -F_{u \tu} (q) & F_{u \tw} (q, \thtast) & 0_{p_c}^T & 0_{p_d}^T  \end{bmatrix}^T$ and \\
$F_e (q, \thtast) = \begin{bmatrix}  F_{e \ty}(q) &   0_{p_b}^T         &   0_{p_f}^T         &- F_{e \te} (q, \thtast) & F_{e \tv} (q, \thtast) \end{bmatrix}^T$.

Note that $\{ u_n \} $ is generated by (\ref{input-ss}) with $\{ s_n \}$ independent of $\{ e_n \}$.
For any nonzero vector $\nu \in \RR^{p_\tht }$, we have
\begin{equation}  \label{nu-e-derivative-ast}
 \nu^T \ole_{\thtast, n} (\tht ) = \nu^T F_u (q, \thtast)u_n + \nu^T F_e (q, \thtast) e_n,
\end{equation}
and, by Parseval's formula,
\begin{equation}  \label{E-nu-e-derivative-ast}
\EE [ |\nu^T \ole_{\tht, n} (\thtast) |^2] = \frac{1}{2 \pi} \int_{- \pi}^{\pi} \begin{bmatrix} \nu^T F_{u} (e^{i \om}, \thtast)  & \nu^T F_{e} (e^{i \om}, \thtast)  \end{bmatrix} \begin{bmatrix}  \Psi_u  (e^{i \om})  & 0 \\ 0 &  \sgmaste^2 \end{bmatrix} \begin{bmatrix}  F_{u}^T (e^{i \om}, \thtast) \nu \\  F_{e}^T (e^{i \om}, \thtast) \nu \end{bmatrix}  \dd \om.
\end{equation}
Since Assumption \ref{assumption-coprime} implies that there does not exist  a vector  $\nu \ne 0_{p_\tht}$ such that
$$
\nu^T F_{u} (e^{i \om}, \thtast) = \nu^T F_{e} (e^{i \om}, \thtast)=0
$$
 for almost all $\om$, (\ref{E-nu-e-derivative-ast}) with (\ref{dfn-u-spectrum-strictpositivity}) yields $\EE [|\nu^T \ole_{\thtast, n} (\thtast) |^2 ]>0 $ for any nonzero
vector $\nu \in \RR^{p_\tht }$, or say, $\EE [ \ole_{\tht, n} (\thtast) \ole_{\tht, n}^T(\thtast) ]>0 $.  Since Assumption \ref{assumption-coprime} holds on some neighborhood of $\thtast$ (see Remark \ref{remark-globalID}), it follows the desired result.

\section*{Appendix B. Notations in adaptive system (\ref{dfn-Dtht})-(\ref{rec-algo-resetting-3}) }  \label{Appendix-adaptivesystem}
 $ \Phi_n = [ \Phi_{1,n}^T \,\, \, \Phi_{2,n}^T \,\, \, \Phi_{3,n}^T \,\, \, \Phi_{4,n}^T \,\, \, \Phi_{5,n}^T \,\, \, \Phi_{6,n}^T \,\, \, \Phi_{7,n}^T \,\, \, \Phi_{8,n}^T ]^T \in \RR^{m + p_\tht + n_\xi + 2}$,
$\Phi_{1,n} = z_n$,  $\Phi_{2,n} = \tu_{n-1}$,  $\Phi_{3,n} =
e_n$, $\Phi_{4,n} =
\xi_n$, $\Phi_{5,n} = \ty_{n-1}$,
$ \Phi_{6,n} = \tw_{n-1} $, $ \Phi_{7,n} = \tv_{n-1} $, $ \Phi_{8,n} = \te_{n-1} $, $ \eta_n = [ e_{n+1} \,\,\, s_n ]^T$,

\begin{eqnarray*}
&& A_\Phi (\tht) = \begin{bmatrix}
A_z  (r (\tht)) &     0  &   0  &  0  &   0  & 0  &   0  &   0  \\ %1
       A_{21}   &  A_{22}  &   0  &  0  &   0  & 0  &   0  &   0  \\ %2
          0    &     0  &   0  &  0  &   0  & 0  &   0  &   0  \\ %3
 B_\xi C_z (r (\tht))  &   0    &   K_\xi  &  A_\xi  & 0  &   0  &   0  &   0  \\ %4
          0    &   0    &  A_{53}   & A_{54} & A_{55}  & 0   &   0  &   0    \\ %5
          0    & A_{62}  &  0   &  0  &   0  &  \tF &   0  &   0  \\ %6
          0    & A_{72} & A_{73}  & A_{74} &   A_{75}  & A_{76} &  A_{77}  &   0   \\ %7
          0    &  A_{82}  &   A_{83}  &  A_{84}  &   A_{85}  &  A_{86} &  A_{87} &   \tC          %8
\end{bmatrix}
, \; \; \;
 B_\Phi (\tht) = \begin{bmatrix}
0 & B_z(r(\tht))   \\%1
0 & B_{22} \\ %2
1 & 0 \\ %3
0 & B_\xi D_z (r (\tht)) \\ %4
0 & 0 \\ %5
0 & 0 \\ %6
0 & 0 \\ %7
0 & 0    %8
\end{bmatrix} , \\
&&  Q (\tht, \Phi_{n+1}) = - R_n ^{-1} (\tht) \e_{\tht, n+1} (\tht)
 \left( C_\xi \Phi_{4, n+1} + \Phi_{3, n+1} + \tht^T \e_{\tht, n+1}  \right), \\
&& R_{n+1} (\tht) = \frac{n}{n+1} R_n (\tht) + \frac{1}{n+1} \e_{\tht, n+1} \e_{\tht, n+1}^T
\end{eqnarray*}

$\e_{\tht, n} (\tht) =  \begin{bmatrix} \Phi_{5,n}  \\ -\Phi_{2,n}  \\
\Phi_{6,n} \\ - \Phi_{8,n} \\ \Phi_{7,n} \end{bmatrix} \in \RR^{p_\tht}$, $A_{21} = \begin{bmatrix} C_z (r (\tht))
\\ 0 \\ \vdots \\ 0 \end{bmatrix} \in \RR^{p_b \times m}$, $ A_{22}   =\tI_{u}
= \begin{bmatrix} 0 & 0 \\ I_{ p_b -1} & 0 \end{bmatrix} \in \RR^{ p_b \times p_b }$, $B_{22} = \begin{bmatrix} D_z  (r (\tht))
\\ 0 \\ \vdots \\ 0 \end{bmatrix} \in \RR^{p_b }$,
$A_{53} =\tI^{e}_{y} = \begin{bmatrix} 1 \\ 0 \\ \vdots \\ 0 \end{bmatrix} \in \RR^{p_a }$,
$A_{54} = \begin{bmatrix} C_\xi \\ 0 \\ \vdots \\ 0 \end{bmatrix} \in \RR^{p_a \times n_\xi }$,  $ A_{55}   =\tI_{y}
= \begin{bmatrix} 0 & 0 \\ I_{ p_a -1} & 0 \end{bmatrix} \in \RR^{ p_a \times p_a }$,
 $A_{62}  = \begin{bmatrix}  \tht_B^T \\ 0 \\ \vdots \\ 0 \end{bmatrix} \in \RR^{p_f \times p_b}$,
 $ A_{72}  = \begin{bmatrix} -\tht_B^T \\ 0 \\ \vdots \\ 0 \end{bmatrix} \in \RR^{p_d \times p_b}$,
 $A_{73}= \begin{bmatrix} 1 \\ 0 \\ \vdots \\ 0 \end{bmatrix} \in \RR^{p_d }$,
$A_{74} = \begin{bmatrix} C_\xi \\ 0 \\ \vdots \\ 0 \end{bmatrix} \in \RR^{p_d \times n_\xi }$,
$ A_{75} = \begin{bmatrix} \tht_A^T \\ 0 \\ \vdots \\ 0 \end{bmatrix} \in \RR^{p_d \times p_a }$, $ A_{76} = \begin{bmatrix} \tht_F^T \\ 0 \\ \vdots \\ 0 \end{bmatrix} \in \RR^{p_d \times p_f }$,
$A_{77}  =\tI_v
= \begin{bmatrix} 0 & 0 \\ I_{ p_d -1} & 0 \end{bmatrix} \in \RR^{ p_d \times p_d }$,
$A_{82} = \begin{bmatrix} -\tht_B^T \\ 0 \\ \vdots \\ 0 \end{bmatrix} \in \RR^{p_c \times p_b}$,
 $A_{83} = \begin{bmatrix} 1 \\ 0 \\ \vdots \\ 0 \end{bmatrix} \in \RR^{p_c }$,
$A_{84} = \begin{bmatrix} C_\xi \\ 0 \\ \vdots \\ 0 \end{bmatrix} \in \RR^{p_c \times n_\xi }$, $ A_{86} = \begin{bmatrix} \tht_A^T \\ 0 \\ \vdots \\ 0 \end{bmatrix} \in \RR^{p_c \times p_a}$,
$ A_{86} = \begin{bmatrix} -\tht_F^T \\ 0 \\ \vdots \\ 0 \end{bmatrix} \in \RR^{p_c \times p_f }$
 and $A_{87} = \begin{bmatrix} \tht_D^T \\ 0 \\ \vdots \\ 0 \end{bmatrix} \in \RR^{p_c \times p_d }$.

%\vskip 0.8cm

\section*{Appendix C. Some useful results in literature}   \label{results_in_literature}
{\it Definition C.1:}    %%%\label{dfn-Mbounded}
A random process $\{\bs_n \}_{n \ge 0}$ is $L$-mixing with respect to the $\sgm$-algebras $(\ff_n , \ff_n^+)$, $n \ge 0$, if the following conditions are satisfied:
\begin{itemize}
\item[i)] $\bs_n$ is $\ff_n$ measurable,
\item[ii)] $\bs_n = O_M (1)$,
\item[iii)] $\sum_{t=0}^\infty \gamma_k (t) < \infty$ for all $1 \le k < \infty$, where
$$
\gamma_k (t) = \sup_{ n \ge t} \EE^{1/k } \left[ \big| \bs_n - \EE [ \bs_n | \ff_n^+ ]\big|^k \right], \,\,\, t \ge 0.
$$
\end{itemize}

%Some useful theorems derived from the main results in \cite{gerencser92,gerencser06} and \cite{huang11} are given as follows, which are applied to develop our results in this paper.
Some useful theorems derived from the main results in \cite{gerencser92,gerencser06} are given as follows, which are applied to develop our results in this paper.

%\begin{condition} \label{condition-1}
{\it Condition C.1:}
The noise $\{ \eta_n \}$ in the system (\ref{dfn-Phi}) is a sequence of independent random variables % of zero mean
such that
\begin{equation} \label{dfn-eta-appendix}
 \sup_n \EE [ \exp{(\a_\eta | \eta_n|^2 )} ] < \infty    %%%\eqno{B.1}
\end{equation}
holds for some $\a_\eta  >0$.
%\end{condition}

%\begin{condition} \label{condition-2}
{\it Condition C.2:}
The time-varying system (\ref{dfn-Phi}) is bounded input-bounded output (BIBO) stable.
%\end{condition}

%\begin{condition} \label{condition-3}

%{\it Condition C.3:}
%The ODE (\ref{dfn-ode}) has an asymptotically stable equilibrium point $X_\ast \in {\rm int} D_X $ with $D_X \subset D_\ast$, where $D_X = D_\tht \times D_R$ and $D_\ast$ is the domain of attraction of $X_\ast$. The initial estimate $X_0 $ is in the interior of $D_{X0} = D_0 \times D_R$, where $D_0$ is a compact set defined by (\ref{dfn-D0}).

%\end{condition}

%\begin{condition} \label{condition-4}
{\it Condition C.3}
The families of matrices $A_\Phi ( \tht)$ and $B_\Phi (\tht)$, $\tht \in D_\tht$, are triply continuously differentiable with bounded partial derivatives
up to second order in $D_\tht$.
%\end{condition}

%\begin{condition} \label{condition-5}
%{\it Condition C.4:}
% Let $D_0 \subseteq D $ be a compact truncation domain such that $X_\ast \in
% {\rm int} D_0$. We assume
% the following: {\rm(i)}~There exists a compact convex set $ D_0' \subset
% D$ such that
% \begin{equation}
% X (t,s,\xi) \in D_0' \quad {\it for} ~\xi \in D_0 ~~{\it and} ~
%X (t,s,\xi) \in D  ~~{\it for} ~\xi \in D_0'
% \end{equation}
% for all $t \ge s \ge 0$. In addition\/ $\lim_{t \rightarrow \infty}  X
% (t, s, \xi) = X_\ast$ for $\xi \in D_0',$ and
% \begin{equation}
% \left\Vert {\partial \over \partial \xi}  X (t,s,\xi)\right\Vert \le C_0
% e^{\alpha (s-t)} \label{eqn:ODE-STAB}
% \end{equation}
% with some $C_0 \ge 1, \alpha > 0$ for all $\xi \in D_0'$ and $t \ge s
% \ge 0$.
% \/ {\rm(ii)}~We have an initial estimate $X_0=\xi_0$ such that for all
% $t \ge s \ge 0$ we have $ X (t, s, \xi_0) \in {\rm int} D_{0}$.\/
%
%
% \begin{remark}  \label{remarkD0}
% The conditions on the existence of $D_0'$ can be removed
% if $D$ itself is convex. Note that $D_\tht$ defined by (\ref{dfn-D_tht}) is convex and hence set $D_0'$ is not needed in our paper.
%\end{remark}
{\it Condition C.4:}
Denote by $X (t; \bt, \bX)$ the solution to ODE (\ref{dfn-ode}) for $t \ge \bt \ge 0$ with $X_{\bt} = \bX$. Assume that (\ref{dfn-ode}) has a unique equilibrium point $X_\ast \in {\rm int} D_{X00} $ on $D_X$ and $\bX \in {\rm int} D_{X00}$, where $D_{X00}  \subset {\rm int} D_X$ is a compact convex set that is invariant for (\ref{dfn-ode}) and $\{ X (t; \bt, \bX): t > \bt \ge 0, \bX \in D_{X00} \} \subset {\rm int} D_{X00}$. Moreover, for every $\bX \in D_{X00}$, we have the Lyapunov exponent $-\a < - 1/2$, i.e., there is a constant $\bC_0 > 0$ such that
\begin{equation} \label{dfn-Lya-exp}
\left| \frac{ \partial} {\partial \bX} X(t; \bt, \bX) \right| \le \bC_0 \exp{ (-\a  (t-\bt \,) )}  %%%%%\eqno{B.2}
\end{equation}
for all $t > \bt \ge 0$.

A variant of \cite[Theorem 4.1]{gerencser92} (see also \cite[Theorem 3.3]{gerencser06}) is given as follows

{\it Theorem C.1:} Assume that  Conditions C.1, C.2, C.3 hold, and that Condition C.4 also holds with $\bX = X_0 = (\tht_0, R_0) \in {\rm int} D_{X00}$. Then  $\{ X_n \}$ with $X_n = (\tht_n, R_n)$ computed by the recursive stochastic algorithm (\ref{dfn-Dtht})-(\ref{rec-algo-resetting-3}) satisfies
\begin{equation} \label{OM-appendix}
 X_n - X_\ast = O_M (n^{-1/2}).    %%%\eqno{B.1}
\end{equation}
In particular $X_n\rightarrow X_\ast$ almost surely as $n\rightarrow\infty$.

\section*{Appendix D. Proof of Theorem \ref{thm-convergence}  }  \label{proof_of_convergence}
It is observed that, by (\ref{dfn-en}) and (\ref{dfn-sn}), Condition C.1 is satisfied.
Let us consider Condition C.2, i.e., the BIBO stability of the linear time-varying
system (\ref{dfn-Phi}).
% which is one of the conditions for convergence of the recursive algorithm (see \cite{gerencser92}, \cite{gerencser06} and \cite{huang11}).
According to Lemma 27.4 in \cite{rugh96}, the time-varying system (\ref{dfn-Phi}) is BIBO stable if the set $\{B_\Phi(\theta)):\theta\in D_\theta\}$ is bounded and the automous system obtained with $B_\Phi(\theta)=0$ is uniformly exponentially stable.

From Appendix B we have that $B_\Phi(\theta)$ depends only on $\theta$
through $B_z(r(\theta))$ and $D_z(r(\theta))$, which by assumption are bounded.

Uniform exponential stability is equivalent to uniform asymptotic
stability (see, e.g., \cite{barabanov05}), which in turn is
equivalent to that the joint spectral radius of the
set of state transition matrices $\Sigma_\Phi = \{ A_\Phi (\tht):\tht
\in
D_\tht \}$ is less than one, when this set is bounded \cite[Corollary 1.1,
  p.21]{jungers09} (see also \cite{berger92}).
Below we will show that $\rho (\Sigma_\Phi)<1$ but first we need to
establish that $\Sigma_\Phi$ is bounded. From Appendix B we have that
$A_\Phi (\tht)$ depends affinely on $A_z(r(\tht))$, $C_z(r(\tht))$,
and $\theta$. There are no other $\tht$-dependencies in $A_\Phi (\tht)$.
By Assumption \ref{assumption-input-signal} $A_z(r(\tht))$,
$C_z(r(\tht))$ are bounded on $D_\theta$. Furthermore, $D_\theta$ is
compact due to Assumption \ref{ass:jointstab}, and hence $\theta\in
D_\theta$ is bounded. Hence $\Sigma_\Phi$ is bounded.
We will now analyze the joint spectral radius of $\Sigma_\Phi$.

%Clearly, the BIBO stability is guaranteed by
%the joint stability of $A_\Phi (\tht)$ for all $\tht \in D_\tht$,
%that is, there exist a symmetric positive definite matrix $V_\Phi$
%and a constant $\lmd_\Phi \in (0,1)$ such that
%\begin{equation}  \label{jointstable-A_Phi}
%A_\Phi^T (\tht) V_\Phi A_\Phi (\tht) \le \lmd_\Phi V_\Phi
%\end{equation}
%for all $\tht \in D_\tht$ (see \cite[Condition 4.1]{gerencser92} and \cite[Cond%%%%%%ition 3.7]{gerencser06}). But
%the BIBO stability of system (\ref{dfn-Phi}) is also ensured when
%the bounded set of matrices $\Sigma_\Phi = \{ A_\Phi (\tht):\tht \in
%D_\tht \}$ is LCP (left convergent products), i.e., every
%left-infinite product $\lim_{n \to \infty} A_n \cdots A_2 A_1$
%converges, where $A_k \in \Sigma_\Phi$ for all $ k = 1, 2, \cdots,
%n$ (see \cite{berger92} and \cite{jungers09}). In this work, we will
%show the BIBO stability of system (\ref{dfn-Phi}) by applying an
%important result of the joint spectral radius (see
%\cite{daubechies92}, \cite{berger92},  \cite{jungers09} and the
%references therein).
Let $ \Sigma_{\Phi}^n = \{ A_n \cdots A_2 A_1: A_k \in \Sigma_{\Phi}, k =1, 2, \cdots, n \}$.
It is easy to observe that every product $\bA_{\Phi, n} \in \Sigma_{\Phi}^n$ is a lower triangular matrix of the form
\begin{equation}  \label{dfn-barA-Phi}
\bar{A}_{\Phi, n} =  \begin{bmatrix}
  \bA_{z, n}   &     0  &   0  &  0  &   0  & 0  &   0  &   0  \\ %1
       \ast   & ( \tI_{u}  )^n &   0  &  0  &   0  & 0  &   0  &   0  \\ %2
          0    &     0  &   0  &  0  &   0  & 0  &   0  &   0  \\ %3
  \ast  &  \ast   &  \ast  &  (A_\xi)^n  & 0  &   0  &   0  &   0  \\ %4
  \ast  &  \ast   &  \ast  &    \ast     &  ( \tI_{y} )^n  & 0   &   0  &   0    \\ %5
  \ast  &  \ast   &  \ast  &    \ast     & \ast  &  \bF_n  &   0  &   0  \\ %6
  \ast  &  \ast   &  \ast  &    \ast     & \ast  &   \ast  &  ( \tI_{v} )^n  &   0   \\ %7
  \ast  &  \ast   &  \ast  &    \ast     & \ast  &   \ast  &   \ast     &   \bC_n          %8
\end{bmatrix}
\end{equation}
for all $n \ge 1$, where $\bA_{z,n} \in \Sigma_z^n$, $\bC_{ n} \in
\Sigma_C^n$, $\bF_{ n} \in \Sigma_F^n$ and the entries denoted by
$\ast$ can be zero or nonzero. Thus the eigenvalues of $\bar{A}_{\Phi,
  n}$ are given by the eigenvalues of the matrices on the block-diagonal.
Obviously, $ ( \tI_{u} )^n$, $ ( \tI_{y} )^n$ and $ ( \tI_{v} )^n$
are strictly lower triangular matrices (i.e., lower triangular
matrices having zeros along their main diagonals) for all $n \ge
1$. In fact, there is a positive integer $n_0$ such that $ ( \tI_{u}
)^n =0$, $ ( \tI_{y} )^n=0$ and $ (\tI_{v} )^n=0$ for all $n \ge n_0$
since all $  \tI_{u}  $, $  \tI_{y} $ and $ \tI_{v} $ are   nilpotent
matrices. Thus these matrices have all eigenvalues at the origin.
Next, note that the transition matrix $A_\xi$ has all its eigenvalues
strictly inside the unit circle, that is $\rho(A_\xi)<1$.
Recalling \eqref{joint-spectral-radius-CF}, the above
gives
\begin{equation}
%\rho_n (\Sigma_\Phi) = \sup \{ [ \rho (\bA) ]^{1/n} : \bA \in
  %\Sigma_{\Phi}^n \} = \max \left\{ \rho_n (\Sigma_z),
\rho_n (\Sigma_\Phi) = \max \left\{ \rho_n (\Sigma_z),
\rho (A_\xi), \rho_n (\Sigma_C), \rho_n (\Sigma_F)\right\}
\end{equation}
for all $n \ge 1$.
But this combined with Assumptions \ref{ass:jointstab} and
\ref{assumption-input-signal} and that $\rho(A_\xi)<1$,
immediately implies
\begin{equation}   \label{joint-spectral-radius-Phi}
\rho (\Sigma_\Phi) = \limsup_{ n \to \infty}  \rho_n (\Sigma_\Phi) < 1 .
\end{equation}
Thus, as argued above, Lemma 27.4 in \cite{rugh96} and \cite[Corollary
  1.1, p21]{jungers09} (see also \cite{berger92}) imply that the switching system (\ref{dfn-Phi}) is (uniformly asymptotically) stable and therefore BIBO stable. Therefore, Condition C.2 is satisfied.

Let us proceed to show the asymptotic stability of the associated ODE (\ref{dfn-ode}). Since  $G (\tht)$ is continuous on the compact $ D_\tht$ (see Appendix A), there exists $\kp_0>0$ such that  $0 \le G (\tht) < \kp_0 I_{p_\tht}$ for all $\tht \in D_\tht$. Note that ODE (\ref{dfn-ode-R}) with initial value $R_0 >0$ gives
\begin{equation} \label{ode-Rt}
R_t = e^{- \frac{1}{2} I_{p_\tht} t} R_0 e^{- \frac{1}{2} I_{p_\tht} t} + e^{- \frac{1}{2} I_{p_\tht} t} \left[ \int_0^t  e^{ \frac{1}{2} I_{p_\tht} \tau}
G(\tht_\tau) e^{ \frac{1}{2} I_{p_\tht} \tau} \dd \tau \right] e^{- \frac{1}{2} I_{p_\tht} t}
\end{equation}
for all $t \ge 0$, which yields
\begin{eqnarray} \label{ode-Rt-bounds}
\kp_r e^{-t}  I_{p_\tht} < e^{- \frac{1}{2} I_{p_\tht} t} R_0 e^{- \frac{1}{2} I_{p_\tht} t} \le R_t \le   R_0  +  e^{- \frac{1}{2} I_{p_\tht} t} \left[ \int_0^t  e^{ \frac{1}{2} I_{p_\tht} \tau}
\kp_0 I_{p_\tht} e^{ \frac{1}{2} I_{p_\tht} \tau} \dd \tau \right] e^{- \frac{1}{2} I_{p_\tht} t} \nonumber \\
 = R_0 + \kp_0 (1- e^{-t}) I_{p_\tht} < \kp_2 I_{p_\tht}  \qquad \qquad \qquad \quad
\end{eqnarray}
for all $t \in [0, \infty)$, where $\kp_2 = \kp_{R}+ \kp_0 $ with $\kp_{R}  I_{p_\tht} > R_0 > \kp_{r}  I_{p_\tht} >0$. This implies that
\begin{equation}  \label{ode-inverseRt-bounds}
\kp_2^{-1} I_{p_\tht} < R_t^{-1} < \kp_r^{-1} e^{t} I_{p_\tht} , \qquad \forall \; t \in [0 , \infty).
\end{equation}
%for all $0 \le t < \infty$.
%
Recall that the asymptotic cost function $W(\tht)$ defined by (\ref{dfn-asym-cost-funct})  has exactly one minimum $\thtast$ on $  D_\tht$ since (\ref{normal-equation}) has the unique solution $\tht=\thtast$. Obviously, $W(\tht) \ge W(\thtast) >0$ for all $\tht \in D_\tht$. By
(\ref{dfn-ode-tht}) and Assumption \ref{assumption-D0}, we observe
\begin{equation}   \label{dW-dt}
\frac{ \dd }{ \dd t } W (\tht) = - W_\tht^T (\tht) R_t^{-1} W_\tht (\tht) \le - \lmd_m ( R_t^{-1} ) \big| W_\tht (\tht) \big|^2 <- \kp_2^{-1} \big| W_\tht (\tht) \big|^2
\end{equation}
for all $t \in [0, \infty)$, and, particularly,
$$
 \frac{ \dd }{ \dd t } W (\tht) \le -\kp_2^{-1} w_\thtast^2 <0
$$
for all $\tht \in D_\tht \backslash D_\thtast$, where $w_\thtast^2 = \inf_{\tht \in D_\tht \backslash  D_\thtast} \big| W_\tht (\tht) \big|^2$.
 This  implies that
there is a finite positive constant $t_{\thtast} \le  W(\tht_0) / (\kp_2^{-1} w_\thtast^2) $ such that $\tht_t \in D_{\thtast}$ for all $t \ge t_{\thtast}$. By Lemma \ref{G-positive-definite}, there is a positive constant $\kp_\thtast$ such that $G(\tht) \ge \kp_\thtast I_{p_\tht}$ for all $\tht \in D_\thtast$.
This combined with (\ref{ode-Rt}) and (\ref{ode-Rt-bounds})  gives
\begin{eqnarray} \label{ode-Rt-t-thtast}
R_t \ge  R_{t_\thtast} + e^{- \frac{1}{2} I_{p_\tht} t} \left[ \int_{t_\thtast}^t  e^{ \frac{1}{2} I_{p_\tht} \tau}
\kp_\thtast I_{p_\tht} e^{ \frac{1}{2} I_{p_\tht} \tau} \dd \tau \right] e^{- \frac{1}{2} I_{p_\tht} t} %\nonumber \\
  > \kp_r e^{- t_\thtast } I_{p_\tht}  + \kp_\thtast e^{- t_\thtast } I_{p_\tht}   %\qquad \qquad \qquad \qquad \qquad
\end{eqnarray}
for all $t \ge t_\thtast$. But (\ref{ode-Rt-bounds}) and (\ref{ode-Rt-t-thtast}) immediately yield $R_t > \kp_1 I_{p_\tht} $ for all $t\ge 0$, where $\kp_1 = \kp_r  e^{- t_\thtast }>0$. This combined with  (\ref{ode-Rt-bounds}) and  (\ref{ode-inverseRt-bounds}) gives
\begin{equation} \label{ode-Rt-bounds-all-t}
 \kp_1 I_{p_\tht} < R_t <  \kp_2 I_{p_\tht} \;\;
 \; {\rm and} \;\;
 \; \kp_2^{-1} I_{p_\tht} < R_t^{-1} <  \kp_1^{-1} I_{p_\tht} %, \quad \forall \; t \ge 0
\end{equation}
for all $t \ge 0$, where positive constants $\kp_1$ and $\kp_2$ can be used to define $D_R$ in (\ref{dfn-DR}).
So (\ref{dW-dt}) holds for all $t \ge 0$. But, according to  \cite[VIII. Theorem, p66]{LaSalle61}, this implies that the equilibrium $\thtast$ of (\ref{dfn-ode-tht}) is asymptotically stable, which also yields  $R_t \to R^\ast = G(\thtast)$ as $t \to \infty$.
Therefore, the equilibrium $(\thtast, R^\ast)$ of ODE (\ref{dfn-ode}) is asymptotically stable.
%But this with Assumption \ref{assumption-D0} implies that Condition C.3 holds. By Theorem C.1, $\{ (\tht_n, R_n ) \}$ computed by the recursive algorithm (\ref{dfn-Dtht})-(\ref{rec-algo-resetting-3}) converges to $(\thtast, R^\ast)$ $a.s.$ as $n \to \infty$.

%%%%Then, as the proof of  \cite[Theorem 4.2]{gerencser92}, the top Lyapunov exponent can be chosen as $-1+ c$ with any $c >0$ since
Finally, we show (\ref{thm-convergence-OM}) as follows.
Note that Assumption \ref{assumption-input-triply} implies Condition C.3 and the Jacobian matrix of (\ref{dfn-ode}) at $(\thtast, R^\ast)$ has the structure
\begin{equation}  \label{jacobian-ast}
\begin{pmatrix} -I_{p_\tht} & 0 \\ \ast & - I_{p_\tht}\end{pmatrix} ,
\end{equation}
all eigenvalues of which are equal to $-1$. It follows that  Condition C.4 is satisfied with the Lyapunov exponent $-\a=-1+ c$ for any $c>0$ in some invariant neighborhood of $(\thtast, R^\ast)$ (see also proof of \cite[Theorem 4.2]{gerencser92}). Let $D_{\tht, R}$ be a compact convex invariant neighborhood such that $(\thtast, R^\ast) \in {\rm int} D_{\tht, R}$ and Condition C.4 is satisfied with the Lyapunov exponent $-\a < -1/2$. The proof of
\cite[Theorem 3.1]{huang11} shows that there exists a sample dependent finite number $N_{\tht, R}$ such that $\{ (\tht_n, R_n ) \}_{n \ge N_{\tht, R}} \subset {\rm int} D_{\tht, R} $ almost surely. Let us consider the sequence $\{ (\tht_n, R_n ) \}_{n \ge N_{\tht, R}}$. But, by Theorem C.1, $\{ (\tht_n, R_n ) \}_{n \ge N_{\tht, R}} $ satisfies
\begin{equation} \label{OM-proof_n-N}
 \tht_n -\thtast = O_M ((n-N_{\tht, R} )^{-1/2}) \quad {\rm and} \quad R_n -R^\ast = O_M ((n-N_{\tht, R})^{-1/2})    %%%\eqno{B.1}
\end{equation}
a.s. as $n \to \infty$. It is noticed that
\begin{equation}  \label{OM-proof_n2n-N}
n^{1/2} = O_M ((n-N_{\tht, R})^{1/2})
\end{equation}
a.s. as $n \to \infty$ since $\PP \{ N_{\tht, R}< \infty \} =1$.
So, by Cauchy-Schwarz inequality,  (\ref{OM-proof_n-N}) and (\ref{OM-proof_n2n-N}) imply (\ref{thm-convergence-OM}). Almost sure convergence follows from (\ref{thm-convergence-OM}) as noted after Theorem 4.1 in \cite{gerencser92}. This completes the proof.

\section*{Appendix E. Proof of Theorem \ref{thm-asy-normality}  }  \label{proof_of_normality}
Since, according to the proof of Theorem \ref{thm-convergence} (see Appendix D), the switching system (\ref{dfn-Phi}) is BIBO stable and
hence is uniformly exponentially stable, there are $C_\Phi >0$ and
$\lmd_\Phi \in (0,1)$ such that (see \cite{michaletzky02})
\begin{equation}  \label{dfn-hPhi}
| \Phi_n | \le C_\Phi \lmd_\Phi^n | \Phi_0| + \sum_{k=1}^n C_\Phi \lmd_\Phi^k | \eta_{n-k} | =:
\hPhi_n.
\end{equation}
Therefore, $\{ \hPhi_n \}$ and hence $\{ \Phi_n \}$ are $L$-mixing
processes since $\{ e_n \}$, $\{ w_n \}$ and hence $\{ \eta_n \}$
are $L$-mixing processes (see \cite{gerencser89}). It follows that the
process
\begin{equation}   \label{Phi-ast}
 \Del \Phi_n = \Phi_n - \Phi^\ast_n
\end{equation}
is  $L$-mixing, where $\{ \Phi^\ast_n \}$  is generated by
(\ref{dfn-Phi}) with $\tht_n = \thtast$ and $\e_{\tht, n}^\ast
=\e_{\tht, n} (\thtast)  = - \varphi_n (\thtast) =  \begin{bmatrix} \ty_{n-1}^T (\thtast) & -\tu_{n-1}^T (\thtast) &
\tw_{n-1}^T(\thtast) & - \te_{n-1}^T(\thtast)  &
\tv_{n-1}^T(\thtast) \end{bmatrix}^T$. Moreover,
since system (\ref{dfn-Phi}) is uniformly exponentially stable
and, by Theorem  \ref{thm-convergence}, $\tht_n \to \thtast$ a.s. as
$n \to \infty$, and using Assumption \ref{assumption-input-triply}
we have $\Del \Phi_n \to 0$ a.s. as $n \to \infty$. Then the stability of
$A_\Phi (\cdot)$ and the
boundedness of $B_\Phi (\cdot)$ imply that $\Del \Phi_n = o_M
(1)$, i.e., $\Del \Phi_n \to 0$ in $\LL_q$-norm for all $q \ge 1$.
Clearly, this yields that
\begin{equation}   \label{Phi-ast-parts}
  \Del \e_{\tht, n} = \e_{\tht, n} - \e_{\tht, n}^\ast
\end{equation}
is an $L$-mixing process and $\Del \e_{\tht, n} = o_M (1)$ since
\begin{eqnarray*}
\Del \e_{\tht, n}= [ \Del \Phi_{5,n}^T \,\,\, -\Del
\Phi_{2,n}^T\,\,\, \Del \Phi_{6,n}^T \,\,\, -\Del
\Phi_{8,n}^T\,\,\, \Del \Phi_{7,n}^T ] ^T,
\end{eqnarray*}
 where   $\Del \Phi_{k,n} = \Phi_{k,n} - \Phi^\ast_{k,n}$ for $k =1,
2, \cdots, 8$.

Note that $\{ \e_{\tht, n}^\ast \}$ is an $L$-mixing process and
therefore
\begin{equation}  \label{e-thtast-variance}
 \frac{1}{n} \sum_{k=1}^n  \e_{\tht,
k}^\ast ( \e_{\tht, k}^\ast )^T \to R^\ast \quad \text{a.s.}
\end{equation}
and hence in law as $n \to \infty$. Moreover, by Cauchy-Schwarz inequality, we observe
\begin{eqnarray}  \label{e-tht-del-variance-1}
&& \frac{1}{n} \sum_{k=1}^n \big[ \e_{\tht, k}^\ast  \Del \e_{\tht,
k} ^T + \Del \e_{\tht, k} (\e_{\tht, k}^\ast)^T + \Del \e_{\tht, k}
 \Del \e_{\tht, k} ^T \big]  \to 0,   \\
\label{e-tht-del-variance-2}
&& \frac{1}{n} \sum_{k=1}^n   \e_{\tht, k}  \Del \tw_{k-1}^T= \frac{1}{n} \sum_{k=1}^n \big( \e_{\tht, k}^\ast + \Del
\e_{\tht, k} \big) \Del \tw_{k-1}^T \to 0, \\
\label{e-tht-del-variance-3}
&& \frac{1}{n} \sum_{k=1}^n   \e_{\tht, k}  \Del \te_{k-1}^T= \frac{1}{n} \sum_{k=1}^n \big( \e_{\tht, k}^\ast + \Del
\e_{\tht, k} \big)  \Del \te_{k-1}^T \to 0, \\
\label{e-tht-del-variance-4}
&& \frac{1}{n} \sum_{k=1}^n   \e_{\tht, k} \Del \tv_{k-1}^T= \frac{1}{n} \sum_{k=1}^n \big( \e_{\tht, k}^\ast + \Del
\e_{\tht, k} \big)  \Del \tv_{k-1}^T \to 0
\end{eqnarray}
in $\LL_q$ for any $q \ge 1$ and hence in law as $n \to \infty$. But, since both $\{ \e_{\tht, n} \}$ and $\{ \e_{\tht, n}^\ast \}$ are  $L$-mixing processes,
  (\ref{e-thtast-variance}) and (\ref{e-tht-del-variance-1}) give
\begin{equation}   \label{e-tht-variance}
\frac{1}{n} \sum_{k=1}^n  \e_{\tht,
k}   \e_{\tht, k}  ^T \to R^\ast
\end{equation}
in $\LL_q$ for any $q \ge 1$ and hence in law as $n \to \infty$. And the combination of (\ref{error-frozen}), (\ref{normal-equation}) and (\ref{e-tht-del-variance-2})-(\ref{e-tht-del-variance-4}) yields
\begin{eqnarray}
\lefteqn{ \left( \frac{1}{n} \sum_{k=1}^n  \e_{\tht,
k}   \e_{\tht, k}^T \right) (\tht_n - \thtast)  }  \nonumber  \\
&& {} = \frac{1}{n} \sum_{k=1}^n  \e_{\tht,
k}  (-\Del \tw_{k-1}^T \thtastF + \Del \te_{k-1}^T \thtastC - \Del \tv_{k-1}^T \thtastD - e_k ) \nonumber  \\
&& {}  \to - \frac{1}{n} \sum_{k=1}^n  \e_{\tht,
k}   e_k
\to  - \frac{1}{n} \sum_{k=1}^n  ( \e_{\tht,
k}^\ast + \Del
\e_{\tht, k} ) e_k  \to  - \frac{1}{n} \sum_{k=1}^n   \e_{\tht,
k}^\ast   e_k   \label{asympt-normal-1}
\end{eqnarray}
in $\LL_q$ for any $q \ge 1$ and hence in law as $n \to \infty$. But, by a martingale central limit theorem (see, e.g., \cite[Theorem 3.2, p58]{hall1980}), we have
\begin{equation}  \label{asympt-normal-2}
\frac{1}{ \sqrt{n}} \sum_{k=1}^n   \e_{\tht,
k}^\ast   e_k \,\,\, \xrightarrow{\LL} \,\,\, \NN (0_{p_\tht}, \sgmaste^2 R^\ast) \quad {\rm as} \quad n \to \infty.
\end{equation}
Recall that the sequence $\{ \e_{\tht, k} \}_{1 \le k \le n}$ is $\ff_{n-1} $ measurable for all $n \ge 1$, where $\e_{\tht, k}$ is the online version of $\overline{\e}_{\tht, k} $ defined by  (\ref{e-derivative}). So, by the martingale central limit theorem,
%%%% (see, e.g., \cite[Corollary 3.1,p58]{hall1980}),
the combination of (\ref{e-tht-variance}), (\ref{asympt-normal-1}) and (\ref{asympt-normal-2}) yields the desired result (\ref{asy-normal-adaptive}). The proof is complete.

\section*{Appendix F. Proof of Theorem \ref{thm-convergence-adaptive} }

The results follow from Theorem \ref{thm-convergence} and Theorem \ref{thm-asy-normality} if we
can verify Assumptions \ref{assumption-input-signal} and
\ref{assumption-input-triply}.
As noted before Theorem \ref{thm-convergence-adaptive}, $A_z$ can be
kept fix and since the basis functions $\{{\mathcal
  B}_k(q)\}_{k=1}^{p_r}$ are stable its spectral radius is less than
one.

Let us now examine the map from $\theta$ to $r$. Firstly,
\eqref{M2}--\eqref{Mk} imply that $M(\hr,\gamma,\theta)$ is
continuously differentiable of any order with respect to $\theta$ on
$D_\theta$. Secondly, if the (primal) problem (\ref{dfn-min-J})-(\ref{dfn-min-M}) is strictly feasible and bounded
from below, and the solution is unique, Theorem 1 in \cite{freund04}
gives that the solution is differentiable with respect to
perturbations of $M$. The essence of the proof is that the equations
(9) in \cite{freund04} have a non-singular Jacobian and hence that the implicit function
theorem applies. Thus, the result of Theorem 1 in \cite{freund04} can
be extended by noting that the equations in (9) are continuously
differentiable of any order, and hence the implicit function theorem
gives that the the solution is continuously differentiable of any
order with respect to perturbations of $M$ \cite{saaty:64}. In summary, the map from
$\theta$ to $r$, as defined by (\ref{dfn-min-J})-(\ref{dfn-min-M}), is continuously differentiable of any order under the assumptions of the theorem.

Next, we study the map from $r$ to the filter coefficients of
$G_u(q,r)$, i.e. the spectral factorization step. For simplicity of
exposition, we restrict our analysis to the FIR case where ${\mathcal
  B}_1(z)=1/2$ and ${\mathcal
  B}_k(z)=z^{-k+1}$ for $k>1$. The case of general rational stable basis
functions can be handled along the same lines but is more involved.
Consider
\begin{align*}
  G_u(z)&=\sum_{k=1}^{p_r}g_k z^{-(k-1)}\\
  \Phi(z)&=G_u(z)G_u(z^{-1})=\sum_{k=1}^{p_r} r_k(z^{-(k-1)}+z^{(k-1)})
\end{align*}
We will use the implicit function theorem \cite{saaty:64} to prove that the map from
$r=\begin{bmatrix}r_1,&\ldots&,r_{p_r}\end{bmatrix}^T$
to $g=\begin{bmatrix}g_1,&\ldots&,g_{p_r}\end{bmatrix}^T$, defined by $\Phi(z)=G_u(z)G_u(z^{-1})$ is
continuously differentiable of any order. Firstly, the map from  $g$
to $r$ is given by
\begin{align*}
r(g):=\oint_{|z|=1} G_u(z)G_u(z^{-1})\Gamma(z)\frac{dz}{z}
\end{align*}
where here $\Gamma(z)=\begin{bmatrix}1 & z^{-1} & \ldots &
z^{-p_r+1}\end{bmatrix}^T$. Differentiating under the integral sign, the Jacobian of this map is
\begin{align*}
J(g):=\oint_{|z|=1} \left(G_u(z)\Gamma(z)\Gamma^T(z^{-1})+G_u(z^{-1})\Gamma(z)\Gamma^T(z)\right)\frac{dz}{z}
\end{align*}
Let $g\neq0$, let $\alpha=\begin{bmatrix}\alpha_1 & \alpha_2 & \ldots &
\alpha_{p_r}\end{bmatrix}^T$ and let $\alpha(z)=\alpha^T\Gamma(z)$ be the
associated polynomial (in $z^{-1}$) of degree $p_r-1$. Suppose that
$J(g)\alpha=0$ for an $g\neq 0$. This can be expressed
\begin{align}
\label{Jalpha}
J(g)\alpha:=\oint_{|z|=1} (G_u(z)\alpha(z^{-1})+G_u(z^{-1})\alpha(z))\Gamma(z)\frac{dz}{z}=0.
\end{align}
Here $G_u(z)\alpha(z^{-1})+G_u(z^{-1})\alpha(z)$ is a symmetric polynomial in
$z^{-(p_r-1)},\ldots,z^{p_r-1}$. Hence, expression \eqref{Jalpha} implies
that this polynomial is identically zero. If $\alpha\neq 0$,
it must hold that $\alpha(z)=G_u(z)$ since $G_u(z)$ and $G_u(z^{-1})$
are coprime. But then
$G_u(z)\alpha(z^{-1})+G_u(z^{-1})\alpha(z)=2G_u(z)G_u(z^{-1})$ is
non-zero, contradicting our assumption that $g\neq 0$. Hence $\alpha=0$ is the only solution to $J(g)\alpha=0$ and
$J(g)$ is non-singular. Using this and that $r(g)$ is continuously differentiable of
any order, it follows from the implicit function theorem that the map
from $r$ to $g$ is continuously differentiable of
any order.

We have now shown that the maps from $\theta$ to $r$, and from
$r$ to $g$ are continuously differentiable of any order.
Furthermore, as already noted before Algorithm
\ref{algorithm-adaptive}, $A_z$ and $B_z$ are fix, whereas $C_z$ and
$D_z$ depend linearly on the filter coefficients $g$ for the used
controllable form. This implies that the maps from $\theta$ to $A_z$,
$B_z$, $C_z$ and $D_z$ are continuously differentiable of any
order on $D_\theta$. Hence Assumption \ref{assumption-input-triply} is satisfied.
In addition, since $D_\theta$ is compact by Assumption
\ref{ass:jointstab}, it follows also from this observation that
the set $\{A_z(r(\theta)),\; B_z(r(\theta)),\; C_z(r(\theta)),\;
D_z(r(\theta)):\theta\in D_\theta\}$ is bounded.  By Step 3) in
Algorithm \ref{algorithm-adaptive}, the random sequence $\{s_n\}$
satisfies the requirements
of Assumption \ref{assumption-input-signal}. Thus all the requirements of
Assumption \ref{assumption-input-signal} are satisfied.

Finally, with the map from $\theta$ to $g$ being continuous and
$D_\theta$ being compact implies that $\|g(\theta)\|$ is bounded on
$D_\theta$. The convergence of sample correlations of
the type \eqref{samplecov} then follows in exactly the same way as (B.2) in
\cite{gerencser09}. This concludes the proof.

\section*{Appendix G. Proof of Lemma \ref{lem:feas}}

    We start with the positivity condition. With $\tilde{r}=\begin{bmatrix}\hr_{p_r} & \ldots &
    \hr_{2}\end{bmatrix}^T$, we can write the matrix in \eqref{dfn-K-Q} as
\begin{align*}
  &  \begin{bmatrix}
    Q & C_u^T\\
    C_u & 2D
    \end{bmatrix}
    -\begin{bmatrix}
  A_u & B_u\end{bmatrix}^T Q \begin{bmatrix}
    A_u & B_u\end{bmatrix}
    =
    \begin{bmatrix}
    Q & \tilde{r}\\
    \tilde{r}^T & r_1
    \end{bmatrix}
    -
    \begin{bmatrix}
    0 & 0\\
    0 & Q
    \end{bmatrix}
\end{align*}
This is a positive definite matrix if we take $\tilde{r}=0$ and $Q$ to be diagonal with strictly
monotonically increasing elements along the diagonal, and take
$\hr_1$ to be greater than the maximal value of $Q$.

Maintaining $\tilde{r}=0$, \eqref{Fisher} gives
\begin{align}
  \label{Gieq}
      G(\theta)= \hr_1\; G_1(\theta)
\end{align}
Take $\alpha\in\RR^{p_\theta}$ to have unit norm. Then
\begin{align}
    \label{G1ieq}
\alpha^TG_1(\theta)\alpha=\frac{1}{\pi}\int_{-\pi}^{\pi}|\alpha(e^{i\omega},\theta)|^2d\omega>0
\end{align}
where $\alpha(z,\theta)=\alpha^T\Lambda(z,\theta)$ is a stable
rational function. The inequality follows since $|\alpha(e^{i \omega},\theta)|^2$ is positive
and has at most a finite number of zeros on the unit circle. Combining
\eqref{Gieq}--\eqref{G1ieq} gives that the minimum eigenvalue of
$G(\theta)$ can be made as large as desired by picking $\hr_1$ large
enough.

In summary, $\hr_1$ large enough and $\hr_2=\ldots=\hr_{p_r}=0$
ensures feasibility of the constraints in Lemma \ref{lem:feas}, and the lemma has been proven.

%\vfill

\end{document}